\documentclass[usenatbib,usegraphicx]{mn2e}

\begin{document}

%\title{XMM-Newton Observations of Two Relic Radio Galaxies,
%	0917+75 and 1401$-$33}
%\title{Limits on Inverse Compton Emission
%	from Relic Radio Galaxies 0917+75 and 1401$-$33
%	from XMM-Newton Observations}
\title[Magnetic field strength of two relic radio sources]
	{Constraints on the average magnetic field strength
	of relic radio sources 0917+75 and 1401$-$33
	from \emph{XMM-Newton} observations}

\author[C. M. H. Chen et al.]
	{C. M. Hubert Chen,$^1$\thanks
		{E-mail: hubert@caltech.edu (CMHC);
			harris@head.cfa.harvard.edu (DEH);
			fiona@srl.caltech.edu (FAH);
			peterm@ess.ucla.edu (PHM)}
	D. E. Harris,$^2$\footnotemark[1]
	Fiona A. Harrison$^1$\footnotemark[1] and
	Peter H. Mao$^1$\footnotemark[1]\thanks
		{Present address:
		Department of Earth and Space Sciences,
		University of California, Los Angeles,
		595 Charles Young Drive East, Box 951567,
		Los Angeles, CA 90095-1567, USA} \\
	$^1$Department of Physics, California Institute of Technology,
		Pasedena, CA 91125, USA \\
	$^2$Harvard-Smithsonian Center for Astrophysics,
		60 Garden Street, Cambridge, MA 02138, USA}
\maketitle

\begin{abstract}
We observed two relic radio sources, 0917+75 and 1401$-$33,
 with the \emph{XMM-Newton} X-ray observatory.
We did not detect any X-ray emission, thermal or non-thermal,
 in excess of the local background level from either target.
This imposes new upper limits on the X-ray flux
 due to inverse Compton scattering of photons
 from the cosmic microwave background by relativistic electrons
 in the relic sources, % (IC/CMB).
 and new lower limits on the magnetic field strength
 from the relative strength of the radio and X-ray emission.
%These upper limits are below the flux levels expected
% from the assumption of equipartition of energy density
% between particles and the magnetic field.
%We discuss the implications of our results with regard
% to physical parameters including the magnetic field,
% the filling factor of the emitting plasma, the proton contribution
% to the particle energy density, and
% possible changes to the spectral slopes of the radio emission
% at low frequencies and of the electron spectrum at low energies.
The combination of radio and X-ray observations provides
 a measure of the magnetic field
 independent of equipartition or minimum energy assumptions.
%While the known population of cluster relics has been growing,
% due to increasing sensitivity in radio observations,
% studies of nonthermal X-ray emission from relics
% remain scarce.
Due to increasing sensitivity of radio observations,
 the known population of cluster relics has been growing;
 however, studies of nonthermal X-ray emission from relics remain scarce.
%Our study adds two relics to the currently small statistics.
%Our study adds two relics to this small X-ray sample.
%Our study adds two relics to the small sample studied in X-rays.
Our study adds to the small sample of relics studied in X-rays.
%In both cases, our field strength lower limits are slightly larger
In both relics, our field strength lower limits are slightly larger
 than estimates of the equipartition magnetic field.
%This result is at odds with recent X-ray field measurements in radio lobes.
\end{abstract}

%\keywords{galaxies: magnetic fields---radiation mechanisms: non-thermal---radio continuum: galaxies---X-rays: galaxies}
\begin{keywords}
galaxies: magnetic fields -- radiation mechanisms: non-thermal -- radio continuum: galaxies -- X-rays: galaxies
\end{keywords}
% Note that the AAS Guide says keywords should be in alphabetical order.
% Other choices
%magnetic fields
%galaxies: jets
%acceleration of particles
% Rejected:
%galaxies: individual (0917+75, 1401$-$33)---

\section{INTRODUCTION}
%---------------------
In a minority of clusters of galaxies,
 there are isolated diffuse sources of radio emission
 with steep spectra and without definite associations with an optical galaxy
 counterpart.
%These sources are classified as haloes or relics, depending on their locations:
These sources are classified as haloes or relics,
 depending on their location and size:
haloes reside at the centre of a cluster of galaxies, and are usually circular
% and not known so far to be highly polarized.
 and not known to be highly polarized.
%In contrast, relics are found at the outskirt of a cluster, and are often
In contrast, relics are found in cluster outskirts, and are often
 elongated and linearly polarized.
%\cite{HarrisMoore}
Given the general phenomenological designation,
 the term `relic' actually refers
 to a few different and probably distinct types of sources,
 including old radio lobes from dead radio galaxies
 and regions where particles are reaccelerated
 in cluster mergers~\citep{Kempner2004,GiovanniniFeretti2004}.
Statistics on relics are growing; to date, over 30~relics have been discovered.
% mostly from low-frequency radio surveys, but also from an X-ray observation
% in one case~\citep{Valtchanov2002}.

The various types of relics share common observational characteristics,
% including a power law radio spectrum, indicating synchrotron emission
 including a power-law radio spectrum, indicative of synchrotron emission
 by relativistic particles in a magnetic field.
%From the radio spectrum alone, the standard approach to determine
% the magnetic field strength and particle density at the source is to assume
The radio spectrum alone constrains only
 the product of the magnetic field and particle density.
The standard approach to estimating
 the magnetic field strength and particle density individually is to assume
 minimum energy or equipartition of energy densities between the two components.
\iffalse
Alternatively, one can determine the magnetic field strength
 by measuring the effect of Faraday rotation on polarization.
However, this approach suffers from large errors,
 and tends to overestimate the field strength
 by factors of a few~\citep*{NewmanNewmanRephaeli2002}.
Yet another approach is by measuring the nonthermal X-ray spectrum
\fi
%An alternative approach is to measure the nonthermal X-ray spectrum
A direct observational approach is to measure the nonthermal X-ray spectrum
 due to inverse Compton scattering of photons from the cosmic microwave
 background (IC/CMB) by relativistic particles
 in the relic~\citep{1979MNRAS.188...25H}.
This provides a measurement of the particle density, and thus decouples
 the contributions of magnetic field and particles to the synchrotron flux.

While IC/CMB measurements have been made on other, stronger radio sources
 of synchrotron emission~\citep[e.g.,][]{Croston2005},
 IC/CMB measurements on relics remain scarce.
The difficulty stems from the weak and diffuse nature of relic emission,
% as well as contamination by thermal emission from the centre of the cluster.
 as well as contamination by thermal emission, often from a nearby cluster.
\iffalse
To date, the only IC/CMB detection in a relic radio source
 is in 0038$-$096 near Abell~85, made with the PSPC on \emph{ROSAT}
 \citep*{1998MNRAS.296L..23B}.  However, the statistics are poor, and
 the observation suffers from thermal contamination.
More recently, \citet*{DurretLimaNetoForman2005} determine from a more
 sensitive observation with \emph{XMM-Newton}
 that the X-ray and radio emissions do not coincide spatially,
% and that the X-ray spectrum does not fit better to a nonthermal model
% than to a thermal model.
 and that the X-ray spectrum fits equally well to a thermal model
 with and without an added power-law component.
\fi
To date, there are only three claims of IC/CMB detections
 in relic radio sources~\citep*{1998MNRAS.296L..23B, Fusco-Femiano2003,
 KempnerSarazinMarkevitch2003}, but the statistics are poor, and later studies
 refute two of these detections~\citep*{DurretLimaNetoForman2005,
 HenriksenHudsonTittley2004}.
For a number of other relics, nondetections have yielded lower limits
 on the magnetic field strength.
% on their magnetic field strength.
%Yet, as \citet{NewmanNewmanRephaeli2002} point out, IC/CMB measurements tend to
% underestimate the field strength.
%Given that IC/CMB emission is an obligatory process in synchrotron sources,
%Thus,
Further X-ray measurements with sensitivity improvement over past studies are interesting,
 as they strengthen the constraints on
 the magnetic field and particle content in relics.

In this paper we present \emph{XMM-Newton} observations of two
relics, 0917+75 and 1401$-$33.
Our X-ray observations are the most sensitive to date for these sources.
Combined with
%low-frequency
 previous
 radio observations, we derive
constraints on their magnetic field and particle content.
% in these sources.
\S\ref{sec:data} describes the X-ray and radio data,
\S\ref{sec:analysis} details our analysis, and
\S\ref{sec:results} presents the numerical results.  We discuss each
source individually in \S\ref{sec:discussion}, and give a summary of
our findings in \S\ref{sec:conclusion}.
We adopt the currently accepted cosmology
 with	$H_0 = 70$\,km~s$^{-1}$Mpc$^{-1}$,
	$q_0 = -0.55$ and
	$\Omega_\Lambda = 0.7$.
At $z=0.125$ and 0.0136, respectively,
 the distances to 0917+75 and 1401$-$33 are 502\,Mpc and 57.8\,Mpc,
 and the linear scales are 134\,kpc/arcmin and 16.7\,kpc/arcmin.

\section{THE OBSERVATIONS}
%=========================
\label{sec:data}
%The targets were observed w/ XMM on date, date and date.
%time-scale.  AOB.
%The relic galaxies we observed are 0917+75, 1401$-$33 and 1253+275.
%Each of the targets---0917+75, 1401$-$33 and 1253+275---were observed
% with \emph{XMM-Newton} in a $\sim 20$\,ksec pointing.
We observed each of the targets, 0917+75 and 1401$-$33,
 in a $\sim 20$\,ks pointing with \emph{XMM-Newton}.
%Radio images of the targets at 1.4\,GHz were obtained from the
We obtained radio images of the targets at 1.4\,GHz from the
 NRAO VLA Sky Survey (NVSS) catalogue \citep{nvss},
 via the \emph{SkyView} facility
% \citep*{skyview}\footnote{\url{http://skyview.gsfc.nasa.gov/}}.
 \citep*{skyview}\footnote{http://skyview.gsfc.nasa.gov/}.
%We also adopted radio parameters from \citet{HarrisSternWillis1993}
We also used published data from \citet{HarrisSternWillis1993}
%, \citet{HarrisWillisDewdney1995},
 and \citet{GossMcadamWellington1987}.
%Parameters of the observation and data are listed in Table~\ref{tab:parameters}.
Table~\ref{tab:source param} lists some properties of the targets
 and the parameters of the observations.
%\input{tab10parameters}
%\input{tab00everything}
%\input{tab02obsParam}
%\placetable{tab:source param}
\begin{table*}
\begin{center}

\caption{Source and XMM observation parameters.
\label{tab:source param}}

\begin{tabular}{rccl}
%\tableline
\hline %\tableline
Target \hfill	& 0917+75	& 1401$-$33 \\ % & 1253+275 \\
\hline %\tableline

Right Ascension (J2000)
		& 09$^h$22$^m$11{\fs}40
				& 14$^h$04$^m$16{\fs}70 \\
						% & 12$^h$55$^m$25{\fs}00 \\
Declination (J2000)
		& 74{\degr}59{\arcmin}31{\farcs}00
				& $-34${\degr}02{\arcmin}22{\farcs}00 \\
					%& $27${\degr}13{\arcmin}46{\farcs}00 \\
Angular Size	& 4\arcmin$\times$8\arcmin
				& 9\arcmin$\times$20\arcmin \\
						%& 10\arcmin$\times$30\arcmin \\
%Galactic \ion{H}{1} Column Density, $n_H$
Galactic \mbox{H\,{\sc i}} Column Density, $n_H$
		& $2.09\times10^{20}$
				& $5.44\times10^{20}$
						% & $8.99\times10^{19}$
								& cm$^{-2}$ \\
Cosmological Redshift, $z$
		& 0.125		& 0.0136	\\ % & 0.023 \\
\hline %\tableline

XMM Observation Date
		& 2002 Mar 20	& 2002 Feb 14	\\ % & 2003 Jun 10 \\
		& 07:00:21.0 UT & 07:59:00.0 UT \\ % & 01:31:00.0 UT \\
XMM Observation Length
		& 26\,744		& 20\,433		% & 23\,700
								& sec \\
\hline %\tableline
\end{tabular}

\end{center}
\end{table*}

%\begin{table}
%\label{tab:parameters}
%\begin{tabular}{lrrr}
%Source			& 0917+75	& 1401$-$33	& 1253+275 \\
%Redshift, $z$		& 0.125		& 0.0136	& 0.023 \\
%Angular size		& $4'\times8'$	& $9'\times20'$ & $10'\times$30'$ \\
%\end{tabular}
%\end{table}

\section{DATA ANALYSIS}
%======================
\label{sec:analysis}
%A quick visual inspection of the images in the \emph{XMM} Pipeline Products
% revealed no detection.
%Thus, additional measures, such as restricting the energy band
% to maximize the signal-to-noise ratio, were necessary.
%Each of the three targets---0917+75, 1401$-$33, and 1253+275---were observed
% with XMM in a 20\,ksec pointing.
We first reprocessed the X-ray data sets using the routines
 \emph{emproc} and \emph{epproc}
 in the \emph{XMM-Newton} Science Analysis Software (SAS), version 6.5.0.
Then, we filtered the processed data based on temporal, spectral,
 and spatial criteria, as described below.

\subsection{Event filtering}
%---------------------------
\label{sec:filtering}
For temporal filtering, we generated good-time intervals (GTIs)
 based on the double-filtering technique
 described in Appendix~A of \citet*{Nevalainen}:
we inspected
 light curves of high-energy events (above 9.5\,keV for MOS and 10.0\,keV for PN)
 of 1-ks resolution,
 and screened out time intervals where the high-energy count rate at any
 of the three detectors exceeded 120 per cent of the mean level.
We also inspected light curves of events within 1--5\,keV
 at the periphery of the field of view
 ($12~\mathrm{arcmin} < r < 15~\mathrm{arcmin}$,
  $r$ being the distance from the nominal pointing direction),
 and screened out time intervals where any of the count rates,
 summed over the periphery annulus, exceeded
 mean+0.028~counts~s$^{-1}$ for MOS and mean+0.056~counts~s$^{-1}$ for PN.
The final GTI for each observation is the intersection of the GTIs
 from the high- and low-energy filters.
It is about 89 per cent the length of the entire observation for 0917+75,
and about 80 per cent for 1401$-$33.
Table~\ref{tab:resultsCalculated} lists the temporal filtering parameters
 in detail.
%\placetable{tab:resultsCalculated}
%We found that only the data set for 1253+275 required flare filtering.
%(See Figs.?)
%We found that the background level was constant in both data sets,
% and flares were absent.
%\footnote{Quantify.}
%Nevertheless, we defined good time intervals based on the criteria described
% by \citet{Nevalainen}.
%\footnote{Did I actually do that for these two sets?}
%\begin{sidewaystable}
\begin{table*}
\begin{minipage}{193mm} %\begin{center}
\renewcommand{\thefootnote}{\thempfootnote}

\caption{Numerical results from the X-ray observations.
\label{tab:resultsCalculated}}

%\tiny
\scriptsize
\begin{tabular}{rr@{}lr@{}lr@{}lr@{}lr@{}lr@{}ll}
%\tableline
\hline %\tableline
Target %\hfill \hspace{0pt}
	& \multicolumn{6}{c}{0917+75}			& \multicolumn{6}{c}{1401$-$33} \\
Detector %\hfill \hspace{0pt}
	& \multicolumn{2}{c}{MOS1} & \multicolumn{2}{c}{MOS2} & \multicolumn{2}{c}{PN}
	& \multicolumn{2}{c}{MOS1} & \multicolumn{2}{c}{MOS2} & \multicolumn{2}{c}{PN} \\
\hline %\tableline

Observation Mode
	& \multicolumn{2}{c}{Full Window} & \multicolumn{2}{c}{Full Window} & \multicolumn{2}{c}{Extended}
	& \multicolumn{2}{c}{Full Window} & \multicolumn{2}{c}{Full Window} & \multicolumn{2}{c}{Extended} \\
	& \multicolumn{2}{c}{} & \multicolumn{2}{c}{} & \multicolumn{2}{c}{Full Window}
	& \multicolumn{2}{c}{} & \multicolumn{2}{c}{} & \multicolumn{2}{c}{Full Window} \\
Filter	& \multicolumn{2}{c}{Medium} & \multicolumn{2}{c}{Medium} & \multicolumn{2}{c}{Medium}
	& \multicolumn{2}{c}{Medium} & \multicolumn{2}{c}{Medium} & \multicolumn{2}{c}{Medium} \\
\hline %\tableline

Hard-band Count Rate
	& \multicolumn{2}{c}{0.188--0.378} & \multicolumn{2}{c}{0.207--0.383} & \multicolumn{2}{c}{0.770--1.65}
	& \multicolumn{2}{c}{0.152--0.208} & \multicolumn{2}{c}{0.151--0.237} & \multicolumn{2}{c}{0.560--0.805} & counts~s$^{-1}$ \\
Mean	& \multicolumn{2}{c}{0.247} & \multicolumn{2}{c}{0.259} & \multicolumn{2}{c}{0.962}
	& \multicolumn{2}{c}{0.184} & \multicolumn{2}{c}{0.188} & \multicolumn{2}{c}{0.671} & counts~s$^{-1}$ \\
Soft-band Annulus Count Rate
	& \multicolumn{2}{c}{0.119--0.202} & \multicolumn{2}{c}{0.118--0.190} & \multicolumn{2}{c}{0.210--0.340}
	& \multicolumn{2}{c}{0.166--0.266} & \multicolumn{2}{c}{0.177--0.251} & \multicolumn{2}{c}{0.368--0.516} & counts~s$^{-1}$ \\
Mean	& \multicolumn{2}{c}{0.148} & \multicolumn{2}{c}{0.150} & \multicolumn{2}{c}{0.262}
	& \multicolumn{2}{c}{0.214} & \multicolumn{2}{c}{0.216} & \multicolumn{2}{c}{0.446}		& counts~s$^{-1}$ \\
Observation Length
	& \,\,\,\,\,\, 27\,825 & & \,\,\,\,\,\, 27\,824 & & \,\,\,\,\,\, 23\,792 &
	& \,\,\,\,\,\, 19\,327 & & \,\,\,\,\,\, 19\,326 & & \,\,\,\,\,\, 15\,291 &		& sec \\
Flare-filtered Length
	& 24\,715 & & 24\,720 & & 20\,693 &		& 15\,483 & & 15\,484 & & 12\,082 &		& sec \\
Good-time Fraction
	& 89\% & & 89\% & & 87\% &			& 80\% & & 80\% & & 79\% \\
Optimal Energy Band
	& \multicolumn{2}{c}{0.2--3.7} & \multicolumn{2}{c}{0.2--3.7} & \multicolumn{2}{c}{0.2--2.3}
	& \multicolumn{2}{c}{0.2--6.5} & \multicolumn{2}{c}{0.2--6.5} & \multicolumn{2}{c}{0.2--4.1}	& keV \\
\hline %\tableline

Source Region \hfill \hspace{0pt} \\
Event Count in This Study, Optimal Band
	& 444 & & 405 & & 1441 &			& 488 & & 529 & & 1244 &			& events \\
	in Blank-sky Data, Optimal Band
	& $\cdots$ & & $\cdots$ & & $\cdots$ &		& 10\,122 & & 9840 & & 24\,308 &		& events \\
	in This Study, Hard Band
	& $\cdots$ & & $\cdots$ & & $\cdots$ &		& 40 & & 42 & & 120 &				& events \\
	in Blank-sky Data, Hard Band
	& $\cdots$ & & $\cdots$ & & $\cdots$ &		& 1403 & & 1432 & & 6698 &			& events \\
Exposure Map,
Median	& 24\,054&.6 & 23\,764&.0 & 17\,680&.7		& 14\,218&.5 & 14\,594&.4 & 10\,030&.4		& sec \\
Mean	& 23\,847&.3 & 23\,061&.4 & 16\,870&.2		& 14\,064&.4 & 14\,295&.8 &    9850&.77		& sec \\
Number of 4.35\arcsec Pixels
	& 1856 & & 1856 & & 1785 &			& 1406 & & 1406	& & 1384 &			& pixels \\
Event Rate\footnote
{Event rates are blank-sky substracted for 1401$-$33
 but not for 0917+75.  See text for details.}
%	& 0.0186 & 0.0176 & 0.0854			& 0.0347 & 0.0370 & 0.126			& counts~s$^{-1}$ \\
	& \multicolumn{2}{c}{0.0186} & \multicolumn{2}{c}{0.0176} & \multicolumn{2}{c}{0.0854}
	& \multicolumn{2}{c}{0.0142} & \multicolumn{2}{c}{0.0168} & \multicolumn{2}{c}{0.0821}		& counts~s$^{-1}$ \\
\hline %\tableline
Background Region \hfill \hspace{0pt} \\
Event Count in This Study, Optimal Band
	& 352 & & 383 &	& 1196 &			& 739 &	& 868 &	& 1840 &			& events \\
	in Blank-sky Data, Optimal Band
	& $\cdots$ & & $\cdots$ & & $\cdots$ &		& 16\,422 & & 17\,797 & & 41\,911 &		& events \\
	in This Study, Hard Band
	& $\cdots$ & & $\cdots$ & & $\cdots$ &		& 63 & & 80 & & 230 &				& events \\
	in Blank-sky Data, Hard Band
	& $\cdots$ & & $\cdots$ & & $\cdots$ &		& 3159 & & 3751 & & 11\,558 &			& events \\
Exposure Map,
Median	& 19\,570&.7 & 19\,612&.4 & 14\,241&.8		& 6043&.56 & 6656&.07 & 4656&.47		& sec \\
Mean	& 19\,268&.2 & 19\,505&.8 & 13\,634&.9		& 5976&.04 & 6701&.31 & 4615&.42		& sec \\
Number of 4.35\arcsec Pixels
	& 1881 & & 1881 & & 1763 &			& 3128 & & 3208 & & 3096 &			& pixels \\
Event Rate\footnotemark[\value{mpfootnote}],
Scaled to Source Region
%	& 0.0180 & 0.0194 & 0.0888			& 0.0556 & 0.0568 & 0.178			& counts~s$^{-1}$ \\
%	& \multicolumn{2}{c}{\,\,\,\,0.0180} & \multicolumn{2}{c}{\,\,\,\,0.0194} & \multicolumn{2}{c}{\,\,\,\,0.0888}
%	& \multicolumn{2}{c}{\,\,\,\,0.0310} & \multicolumn{2}{c}{\,\,\,\,0.0319} & \multicolumn{2}{c}{\,\,\,\,0.0974} \\
	& 0.01&80 & 0.01&94 & 0.08&88			& 0.03&10 & 0.03&19 & 0.09&74 \\
													% & counts~s$^{-1}$ \\
%Scaled to Source Region
%	& \multicolumn{2}{c}{$\pm 0.0016$} & \multicolumn{2}{c}{$\pm 0.0015$} & \multicolumn{2}{c}{$\pm 0.0016$}
%	& \multicolumn{2}{c}{$\pm 0.0043$} & \multicolumn{2}{c}{$\pm 0.0035$} & \multicolumn{2}{c}{$\pm 0.0129$} & counts~s$^{-1}$ \\
	& $\pm$0.00&16 & $\pm$0.00&15 & $\pm$0.00&16	& $\pm$0.00&43 & $\pm$0.00&35 & $\pm$0.01&29	& counts~s$^{-1}$ \\
\hline %\tableline
%Source Count Rate, $2\sigma$ Upper Limit \hfill \hspace{0pt}
%	& $\leq 3.13 \times 10^{-3}$
%	& $\leq 3.06 \times 10^{-3}$
%	& $\leq 3.19 \times 10^{-2}$	& $\leq 8.25 \times 10^{-3}$
%					& $\leq 7.03 \times 10^{-3}$
%					& $\leq 2.98 \times 10^{-2}$ &counts~s$^{-1}$\\
Source Count Rate, $3\sigma$ Upper Limit \hfill \hspace{0pt}
	& \multicolumn{2}{c}{$\leq 4.70 \times 10^{-3}$}
	& \multicolumn{2}{c}{$\leq 4.59 \times 10^{-3}$}
	& \multicolumn{2}{c}{$\leq 4.78 \times 10^{-2}$}	& \multicolumn{2}{c}{$\leq 1.28 \times 10^{-2}$}
								& \multicolumn{2}{c}{$\leq 1.05 \times 10^{-2}$}
								& \multicolumn{2}{c}{$\leq 3.86 \times 10^{-2}$} & counts~s$^{-1}$\\
(MOS Co-added)
%	& \multicolumn{2}{c}{$\leq 2.57 \times 10^{-3}$} &
%	& \multicolumn{2}{c}{$\leq 7.18 \times 10^{-3}$} & & counts~s$^{-1}$ \\
	& \multicolumn{4}{c}{$\leq 3.86 \times 10^{-3}$} & &
							& \multicolumn{4}{c}{$\leq 1.11 \times 10^{-2}$} & &	& counts~s$^{-1}$ \\
\hline %\tableline
\end{tabular}

\end{minipage} %\end{center}
%\end{sidewaystable}
\end{table*}

\begin{figure*}
%\plotone{image0917+75.eps}
%\plotone{f1.eps}
%\includegraphics[width=0.5\textwidth]{f1.eps}
\includegraphics[width=\textwidth]{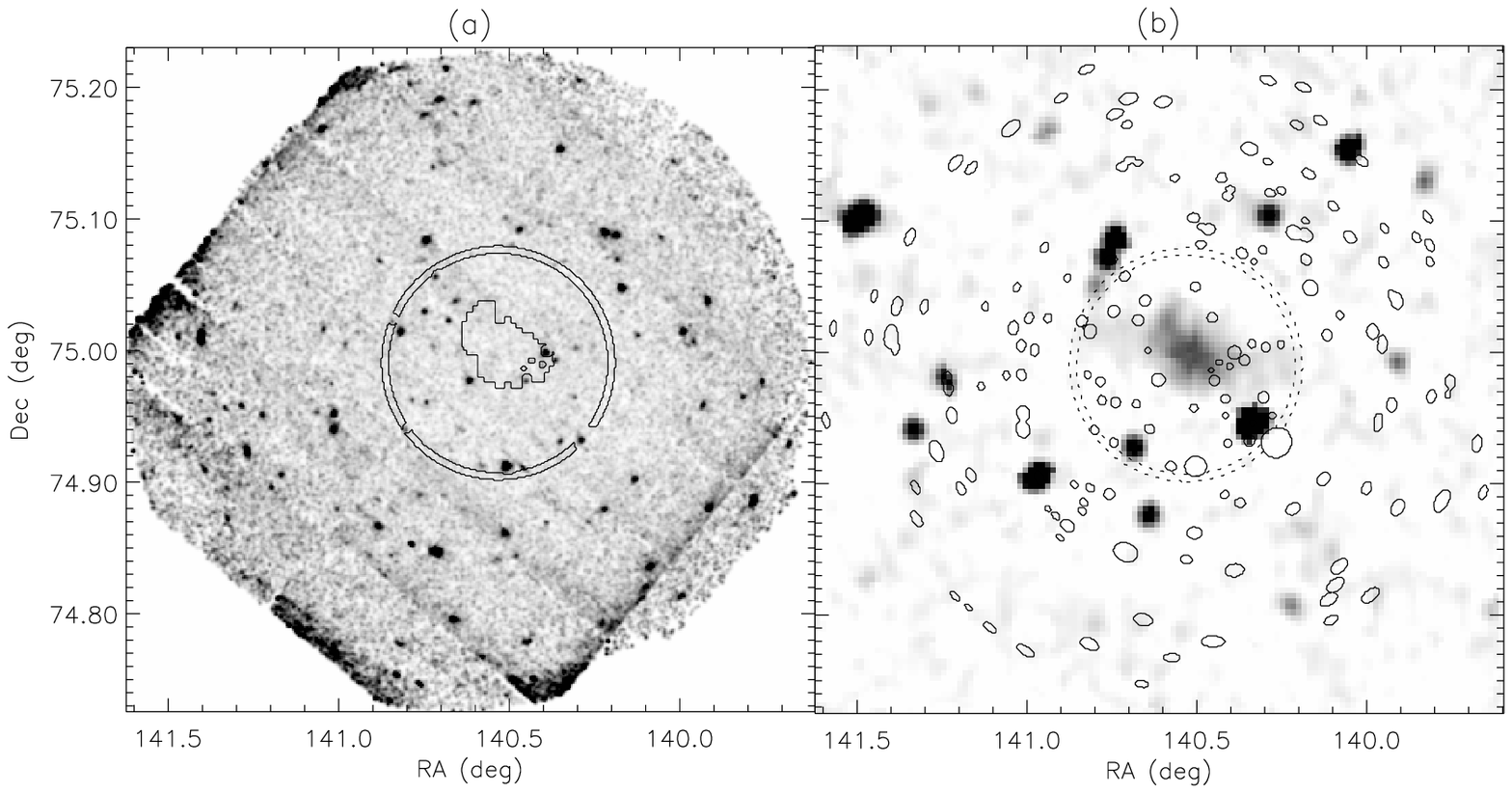}
\caption{\emph{XMM-Newton} (a) and NVSS (b) images of 0917+75.
 The two images have the same spatial extent.
 The source (centre) and background (annulus) regions are
  indicated by solid lines in (a);
  the background region is also shown in (b) by dotted lines.
 X-ray point sources detected within the \emph{XMM-Newton} field of view are
  marked by solid lines in (b), with the size of each region proportional
  to the X-ray flux of the point source.
 The boundary of the source region is
  identical to the mean+5$\sigma$ contour in the NVSS image,
  excluding all overlapping X-ray point source regions.
 Image (a) is made by
  (1) scaling the filtered MOS (0.2--3.7\,keV) and PN (0.2--2.3\,keV) images,
  according to each instrument's response to the same source model and flux,
  (2) correcting the scaled images for exposure and combining them
  with the SAS routine \emph{emosaic}, and
  (3) smoothing the combined image with a Gaussian function
  of width $\sigma = 4$~arcsec,
  about the size of the instrument's point spread function.
\label{fig:0917+75}}
\end{figure*}

\begin{figure*}
%\plotone{image1401-33.eps}
%\plotone{f2.eps}
%\includegraphics[width=0.5\textwidth]{f2.eps}
\includegraphics[width=\textwidth]{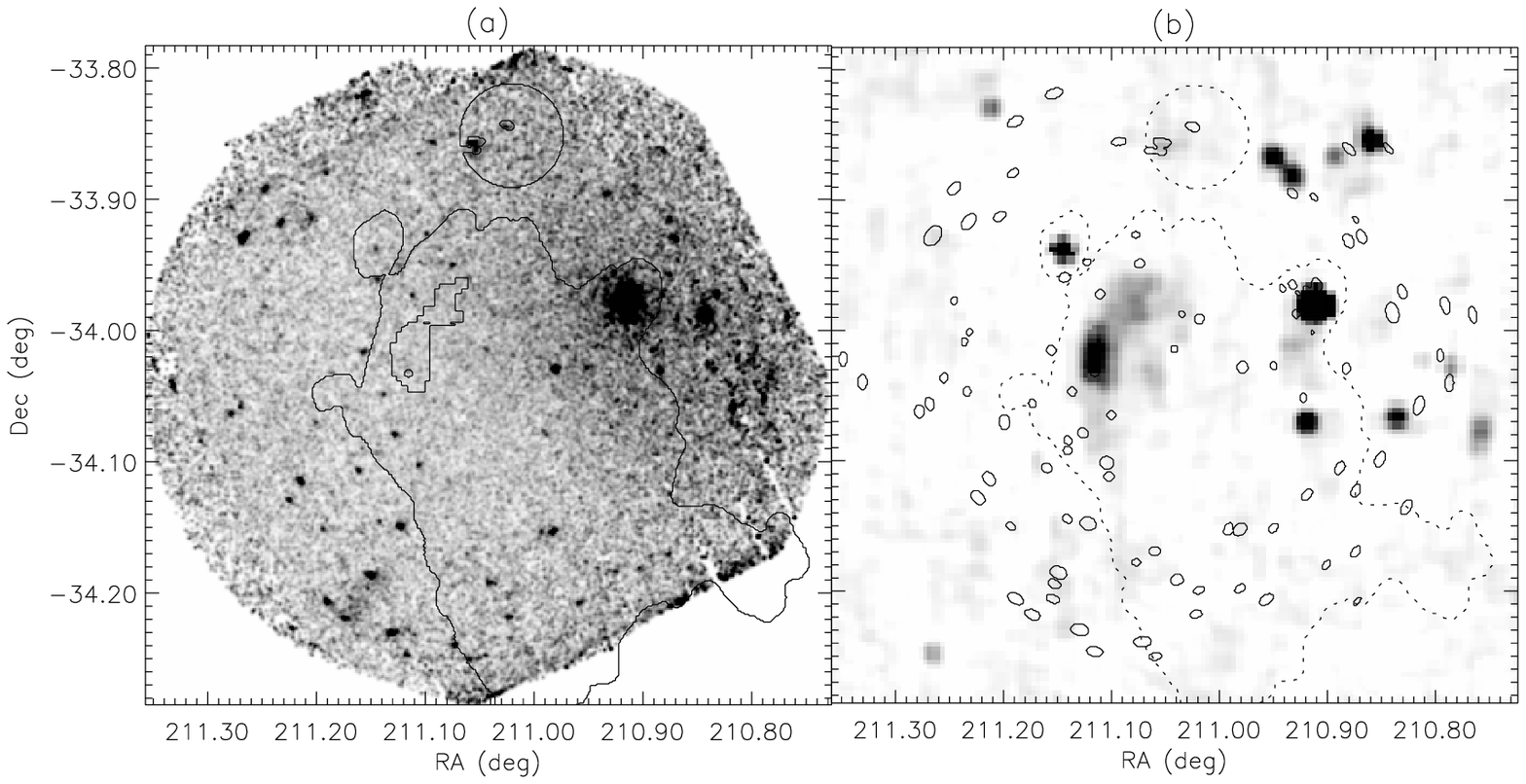}
\caption{\emph{XMM-Newton} (a) and NVSS (b) images of 1401$-$33.
 See Fig.~\ref{fig:0917+75} for a description of the various features.
 In addition to the source and background regions
  at the centre and top of the images, respectively,
  we also mark the extent of the radio emission at 330\,MHz,
  according to fig.~1 in \citet{Subrahmanyan2003}.
 Note that the source and background regions here are equidistant
%  (in the plane of the sky) from the bright galaxy NGC~5419
%  at the centre of the Abell~S753 cluster, seen here in both images
  from the bright galaxy NGC~5419 at the centre of the Abell~S753 cluster,
  about 10~arcmin to the west (right) of the source region.
 Image (a) combines the filtered MOS images of 0.2--6.5\,keV
  and the filtered PN image of 0.2--4.1\,keV.
% Note also that at lower radio frequencies (100s of MHz),
%  the radio emission of 1401$-$33 extends substantially to the southwest.
% See \citet{Subrahmanyan2003} for details.
\label{fig:1401-33}}
\end{figure*}

% to restrict the X-ray flux measurements to within the extent of each target,
% as observed at radio frequencies,
% and to screen out contributions by X-ray point sources that overlap
% with the extended targets.
We restricted the region for photon extraction
 to coincide with the radio source extent,
 and we filtered out regions containing point source detections.
%The purpose of our spatial filtering step was
%Radio image of each target at 1.4\,GHz was obtained from the NVSS catalogue,
% via SkyView (\url{http://skyview.gsfc.nasa.gov}).
From the NVSS radio image of each target, we made a mask
 by selecting only pixels in the vicinity of the target
 whose flux densities are at least 5$\sigma$ above the background level
 in each image.
%\footnote{Actually, is this a problem?  If we underestimate the size
% of the radio emitting region, then we would underestimate the X-ray flux,
% thus overestimate the magnetic field, thus making our results invalid.
% Yet, of course, the spatial extent at this high a frequency is irrelevant
% anyway, at least for 0917+75.}
To mask out X-ray point sources in each field, we took the three source lists
 (one from each detector on \emph{XMM-Newton})
 provided with the Pipeline Products,
 and ran them through the SAS routine \emph{region}.
This produced three FITS region files for each field,
 specifying regions that excluded all the detected point sources,
 with the size of each excluded region determined by the level of flux
 for that particular source.
We combined the radio mask and X-ray point source regions
 to form the spatial filtering expression for each of the targets.
%(See Figs.?)
These masks are shown in Figs.~\ref{fig:0917+75} and~\ref{fig:1401-33}.
%\placefigure{fig:0917+75}
%\placefigure{fig:1401-33}
%\begin{figure}
%\label{fig:0917+75}
%\plotone{image0917+75.eps}
%\caption{XMM-Newton and NVSS images of 0917+75}
%\end{figure}
%\begin{figure}
%\label{fig:1401-33}
%\plotone{image1401-33.eps}
%\caption{XMM-Newton and NVSS images of 1401$-$33}
%\end{figure}

% spectral filtering...
%The objective of spectral filtering was
% to maximize the signal-to-noise ratio, and thus the likelihood of a detection.
%In order to maximize the likelihood of a detection,
We determined the optimal energy band for signal extraction
% that maximizes the signal-to-noise ratio for each source,
 by comparing (1) the expected source flux density at equipartition
% and the average background flux density for \emph{XMM-Newton}
 and (2) the average background for \emph{XMM-Newton}
 at various energies.
%The source flux level was obtained with \emph{PIMMS}, using a power-law model
% for each target, with the parameters listed in Table~\ref{tab:source param}.
To estimate the source flux density at equipartition,
 we followed the analytical formulation in \citet{1979MNRAS.188...25H}
 and deduced a power-law model for the IC/CMB emission from each source,
 with model parameters based on data from past radio observations
 \citep{HarrisSternWillis1993}.
Tables~\ref{tab:source param} and~\ref{tab:resultsInferred} lists
 these parameters.
%\placetable{tab:resultsInferred}
%The background level was obtained from blank sky data sets provided
% on the `Background Analysis' pages at the XMM Science Centre web site.\cite{XMMbackground}
%For the background flux density, we used blank sky event files
To estimate the background, we took blank sky event files
 obtained from the web site of the \emph{XMM-Newton} Science Operation
% Centre\footnote{\url{http://xmm.vilspa.esa.es/external/xmm\_sw\_cal/background/blank\_sky.shtml}} as representative background measurements,
 Centre\footnote{\scriptsize http://xmm.vilspa.esa.es/external/xmm\_sw\_cal/background/blank\_sky.shtml} as representative background measurements,
 and computed a background spectrum from the files.
\citet{CarterRead2007} described these blank sky data sets in detail.
%The expected signal-to-noise ratios were computed for energy bands with
% maximum and minimum energy values at increments of 0.1\,keV.
%The band with the highest signal-to-noise ratio for each source spectrum
% and detector background was selected.
With these source and background spectra, we selected an energy band
 $[E_\mathit{min}, E_\mathit{max}]$ and calculated the signal-to-noise
 ratio (S/N) associated with the band.
We then varied $E_\mathit{min}$ and $E_\mathit{max}$ at 0.1-keV increments
 until we obtained the energy band with the maximum S/N.
Using this method, we arrived at the optimal MOS bands
 0.2--3.7\,keV for 0917+75 and 0.2--6.5\,keV for 1401$-$33.
Table~\ref{tab:resultsCalculated} shows the selected energy bands
 for PN as well.

%In addition to the GTIs and energy bands obtained from above,
In addition to the GTIs and energy bands obtained as described above,
 we further applied the following event filtering expressions:
% (\verb$#XMMEA_EM && PATTERN <= 4$ for MOS
% and \verb$#XMMEA_EP && PATTERN <= 12$ for PN).
% (\verb$FLAG == 0 && PATTERN <= 12$) for MOS
% and (\verb$FLAG == 0 && PATTERN <= 4$) for PN.
 (\texttt{FLAG == 0 \&\& PATTERN <= 12}) for MOS
 and (\texttt{FLAG == 0 \&\& PATTERN <= 4}) for PN.
These expressions retained only events triggering 1 or 2 pixels,
 and only those flagged as valid by the standard processing routines.
From the filtered event list, we generated images and exposure maps.
We then used the masks and regions described above to select
 the region of sky corresponding to each target.
%The photon count rate is obtained as the quotient of the number of counts
% in the selected region divided by the sum of the corresponding
% exposure map values.
The total counts divided by the mean of the exposure map in the selected region
 gave the event rate.
% gave the photon count rate;

%\section{RESULTS}
%-----------------
%\label{sec:results}
\iffalse
\footnote{Per Fiona's advice, I substantially reorganised this section.
The outline in typewriter text above is just FYI, and will be removed
 before submitting.}
\scriptsize
\begin{verbatim}
%1. Spell out numerical results.  Conclusion: No detection.
%2. Constaints: How to get background sigma; spell out sigma; compare w/ src;
%		B limit; conclusion: B is higher than equipartition.
%3. 0917+75 background: a. ACE data; b. blank sky data; c. light curves.
%		Conclusion: clueless, but non-detection OK, affects limits.
%4. 1401-33 background: a. thermal emission; b. vignetting
%		Conclusion: imperfect, but we've done all we can.
\end{verbatim}
\normalsize
\fi
\subsection{Background determination and flux upper limit analysis}
%------------------------------------------------------------------
\label{sec:background}
%The background measurements for both observations require additional discussion.
The two observations in this paper pose very different challenges
 for background determination.  We discuss our approach for each observation
 seperately in the following.

\subsubsection{0917+75}
\begin{figure}
%\plotone{ace.eps}
%\plotone{f3.eps}
%\includegraphics[width=0.5\textwidth]{f3.eps}
\includegraphics[width=0.47\textwidth]{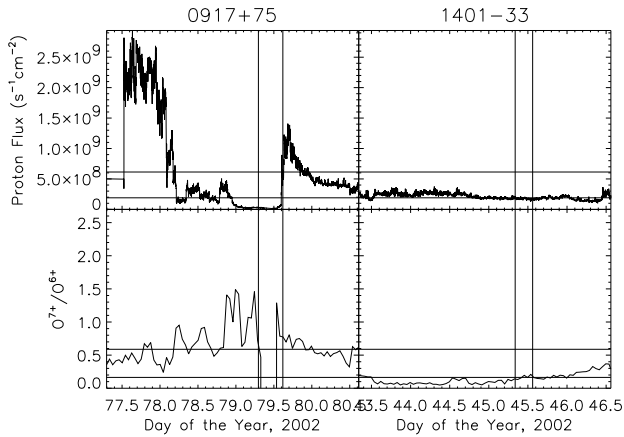}
\caption{Proton flux and ratio of oxygen charge states O$^{7+}$/O$^{6+}$
  in the solar wind, measured by \emph{ACE} during the time of our observations.
 The two horizontal lines in each panel indicate the mean and 90-percentile
  levels of each quantity over the first 100~days of 2002.
 The two vertical lines in each panel delimit
  the period of the \emph{XMM-Newton} observation.
 The 64~s-averaged proton flux data are
  from the SWEPAM instrument on \emph{ACE},
  while the hourly charge state ratios are from the SWICS/SWIMS instrument.
 Note that the break in the O$^{7+}$/O$^{6+}$ curve
  at the time of our observation of 0917+75 was due to missing data.
 Yet, from the values before and after the break,
  one can infer that the O$^{7+}$/O$^{6+}$ ratio during the observation
  is probably higher than the 90-percentile.
 There is a $\sim$1~hour travel time by the solar wind from the L1 point,
  the location of \emph{ACE}, to the Earth.
 The time shown here is the time measured by \emph{ACE}.
\label{fig:ace}}
\end{figure}

%^^^^^^^^^^^^^^^^^^^^^^
For 0917+75, the time of our observation unfortunately coincided
 with a coronal mass ejection (CME) from the Sun.
\citet*{Snowden} reported an enhancement in the \emph{XMM-Newton} background,
 especially at 0.5--1.0\,keV,
 concurrent with an enhancement in the solar wind
 measured by the \emph{Advanced Composition Explorer} (\emph{ACE})
 and other monitoring spacecraft.
When we obtained data
% from \emph{ACE}\footnote{\url{http://www.srl.caltech.edu/ace/ASC/}}
 from \emph{ACE}\footnote{http://www.srl.caltech.edu/ace/ASC/}
 and inspected the light curves 
 at the time of our observations (Fig.~\ref{fig:ace}),
 we found that our observation of 0917+75 coincided with an episode
 of significant enhancement of the O$^{7+}$/O$^{6+}$ ratio in the solar wind,
 characteristic of a CME (R.~C.~Ogliore 2006, private communication).

\begin{figure}
%\plotone{spectrum0917+75.eps}
%\plotone{f4.eps}
%\includegraphics[width=0.5\textwidth]{f4.eps}
\includegraphics[width=0.47\textwidth]{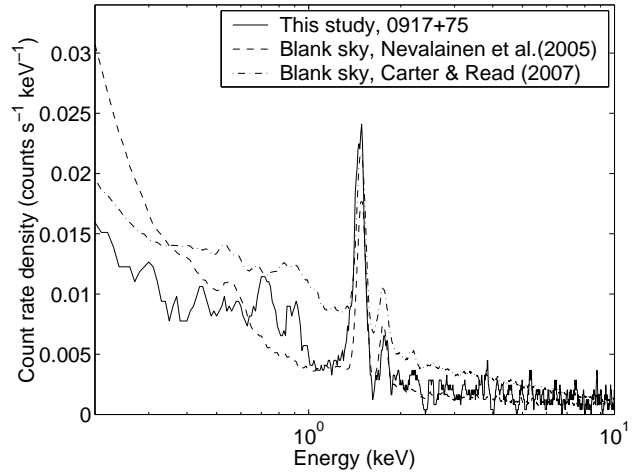}
\caption{Spectra of 0917+75 and of the blank sky data.
 The solid line shows the spectrum of 0917+75 from MOS2,
  convolved with a 100~eV-wide boxcar.
 The dashed and dash-dot lines show the blank sky spectra
  (also convolved with a 100~eV-wide boxcar)
  from \citet{Nevalainen} and from \citet{CarterRead2007}, respectively.
 Both blank sky spectra contain events
  extracted from the same detector coordinates of MOS2
  as the source region in this study,
  and scaled to the spectrum of 0917+75
  by equating the GTI-filtered livetimes % length of the GTIs and livetimes.
 Data from \citet{CarterRead2007} is also refiltered with (\texttt{FLAG == 0})
  for consistent comparison.
%  and the count rates above 9.5~keV, respectively,
  in the data sets.
 Note the excess of counts at discrete `spectral lines' at 0.5--1.0~keV
  when compared to the data from \citet{Nevalainen},
  and the overall deficit of counts when compared to \citet{CarterRead2007}
% Note the deficit of counts in the source spectrum,
%  compared to the blank sky spectrum in either normalization scheme.
% Note the elevated count rate in the source spectrum at 0.5--1.0~keV
  (see text for an explanation).
% The count rates displayed here are not corrected for dead time.
 The count rates displayed here are livetime corrected.
\label{fig:0917+75spectrum}}
\end{figure}

To assess the impact of this event on the background level in our observation,
% we repeated the background measurement using two sets
% of blank sky event files---from \citet{Nevalainen} and \citet{CarterRead2007},
% respectively---which serve as independent, standard data sets.
 we compared the event rate in our data set with the rates
 in two standard blank-sky data sets independent of our observation:
 one set was from \citet{Nevalainen} and the other from \citet{CarterRead2007}.
% that we used for spectral filtering (see \S\ref{sec:filtering}),
We chose these two sets because we employed the flare-filtering recipe
 of \citet{Nevalainen} in this study, while \citet{CarterRead2007} provides
 a much larger data set, with better event statistics.
We refiltered the \citeauthor{CarterRead2007} data set
 with the (\texttt{FLAG == 0}) expression
 so that comparison of the three data sets is consistent.
However, we have not refiltered the \citeauthor{CarterRead2007} data set
 for flares using the recipe in \citet{Nevalainen},
 as its full-view hard-band event rates
 (0.26 and 2.7~counts~s$^{-1}$ for MOS and PN, respectively)
 are above the range deemed consistent with the sample of \citet{Nevalainen},
 making the hard-band filtering method inapplicable.
To minimize the variability of any other systematic parameters
 (such as vignetting),
 we applied the masks and regions that we obtained
 for the source regions of 0917+75 in this blank-sky measurement,
 thus selecting the same detector pixels.
We computed count rates at various energies
 from the filtered blank sky event files,
 and compared them with our source-region count rates for 0917+75,
 both without exposure map correction.
Fig.~\ref{fig:0917+75spectrum} shows spectra of 0917+75
 and of the blank sky data thus obtained.
%It shows that the continuum level is lower for 0917+75
% than for the blank sky data.
%The result is against our expectation---despite evidence of a CME
% from the \emph{ACE} data, the source event rates for 0917+75 are
% still lower than the blank sky rates from the same regions of the detectors.
It shows that the continuum level in our observation of 0917+75 is consistent
 with that from \citet{Nevalainen}, but lower than the level from
 \citet{CarterRead2007}.
We believe the difference between the two blank sky data sets is
 due to different levels of flare filtering.
As we employed the more stringent double-GTI filtering recipe
 of \citet{Nevalainen} in this study, our spectrum of 0917+75 should be
 directly comparable with a spectrum of the same detector region made
 from their data set.
We note that the CME during our observation explains the excess line emission
 (relative to the continuum) at multiple energies below~1.0\,keV
 in the spectrum of 0917+75, which is not seen in either blank sky spectrum.

\begin{figure*}
%\plotone{lightcurves0917+75.eps}
%\plotone{f5.eps}
\includegraphics[width=\textwidth]{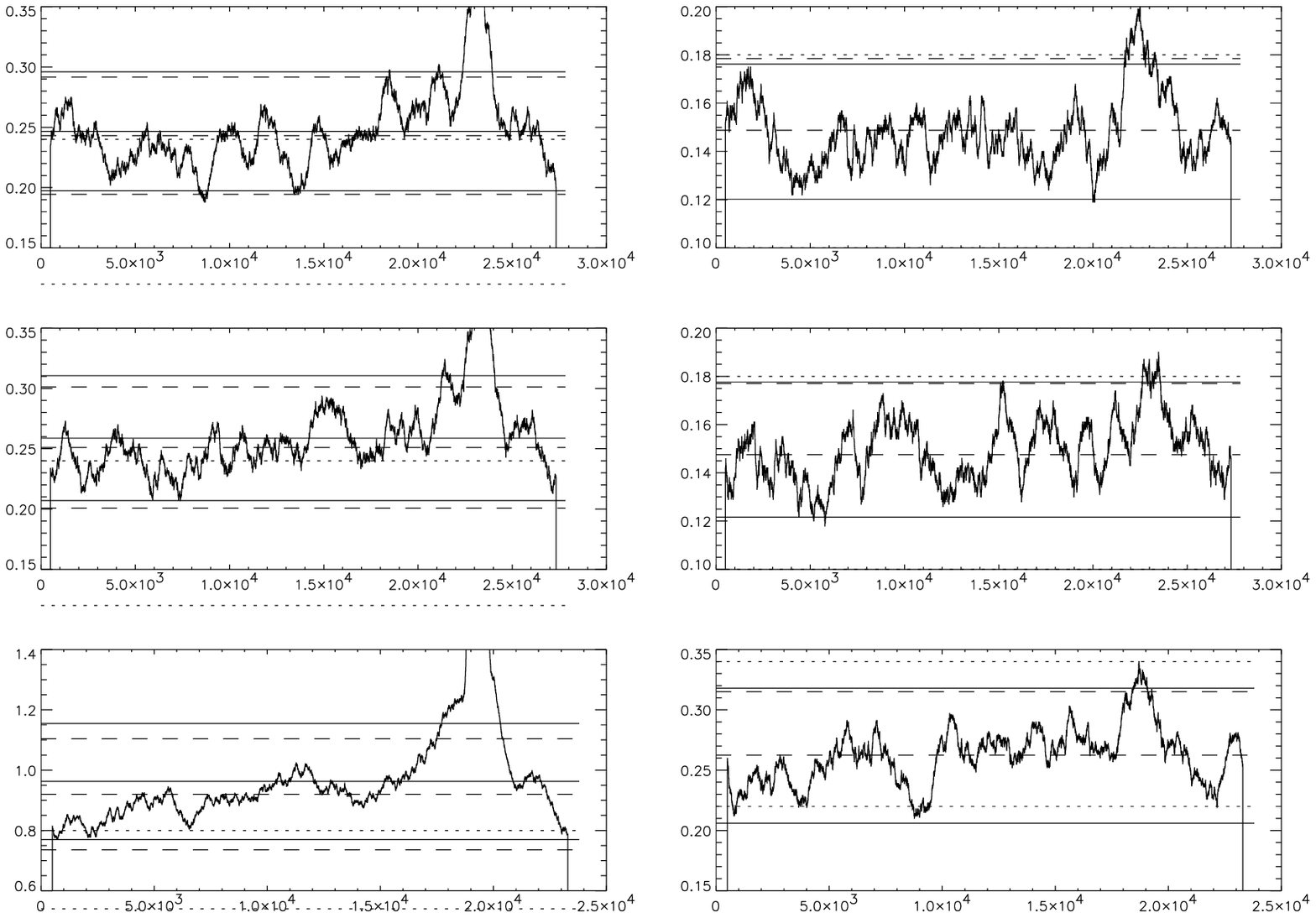}
\caption{Light curves of high-energy events (left)
  and of events at the periphery of the field of view (right)
  during the observation of 0917+75.
 The three rows of graphs show, from top to bottom,
  events from MOS1, MOS2 and PN, respectively.
 The abscissae show the time since the start of the observation, in seconds.
 The ordinates show the count rate, in counts~s$^{-1}$.
 The three solid lines in each graph on the left indicate
  the mean and (mean$\pm$20\%) high-energy count rates.
 The long-dash lines show the same quantities with the median.
 The two short-dash lines delimit the `normal' range of count rates
  reported by \citet{Nevalainen},
  which is exceeded during most of this observation.
 On the right, the solid (long-dash) lines delimit the acceptable range
  of count rates computed with the mean (median),
  according to \citet{Nevalainen}.
 The short-dash lines delimit the `normal' range
  observed by \citet{Nevalainen};
  only the top line is seen in the top two graphs.
\label{fig:0917+75light curve}}
\end{figure*}

To further investigate this background discrepancy, we looked at light curves
 generated from our observation of 0917+75.
Fig.~\ref{fig:0917+75light curve} shows the light curves of all events
 above about 10\,keV from the entire field of view, which we generated
 to determine the high-energy GTIs (see \S\ref{sec:filtering}).
%Although the count rates within the optimal energy band of 0.2--3.7\,keV were
% lower for 0917+75 than for and the blank sky data,
This figure shows that
% the count rates over the entire field of view above about 10\,keV,
% as shown in these light curves, were above the `acceptable' range
 the count rates over the entire field of view above about 10\,keV
 were above the `acceptable' range
 of high-energy particle background rate reported by \citet{Nevalainen}
 for the majority of our observation period, and for all three detectors.
This is evidence that our entire observation was
 plagued with low-amplitude long-duration flares,
 and that residual flares still lurk in the background
 after our double-GTI filtering.
Thus, these light curves are consistent with the \emph{ACE} data,
 suggesting an elevated background.

Because our observation was contaminated by flares,
 count rates from blank-sky data sets do not correctly reflect
 the nature of the background in our data.
Besides, according to Fig.~\ref{fig:0917+75spectrum},
 the source rate in our data does not exceed the blank-sky count rates.
% in other words, we have a non-detection.
%Thus, in addition to the background level itself,
% we need to measure its \emph{variation} in order to set a flux limit
% we need to measure its variation in order to set a flux limit
% on 0917+75.
%For these two reasons, we considered multiple local background regions.
Both of these reasons make it impossible for us to use blank-sky data
 for quantitative background subtraction.
Thus, we measured the background level and its variation
 (for setting a flux limit in the case of non-detection)
 solely from multiple local background regions.
We obtained the local background regions using a similar approach as we did
 for the source regions,
 applying temporal and spectral filters
 described in \S\ref{sec:filtering}.
For the spatial filters, we inspected the radio image of 0917+75 from NVSS,
% of each \emph{XMM-Newton} pointing,
 as well as X-ray images from the Pipeline Products,
 to find regions of sky within the \emph{XMM-Newton} field of view
% where the radio flux is low,
 but securely outside the target, as seen in radio frequencies,
 and where there are few detectable X-ray point sources.
%We first drew these regions by hand, using the polygon region tool in ds9,
% over the radio and X-ray images.
%For 0917+75, we selected two background regions:
%The first one was an annulus immediately outside the target,
%For 0917+75,
We selected an annulus immediately outside the target,
 with area similar to the source region.
This annulus is inscribed within the central CCDs of MOS1 and MOS2,
% which simplifies background subtraction,
 which simplifies the comparison between the source and background regions,
 as potentially different background levels at different CCDs
 \citep{PradasKerp2005} become irrelevant for the two MOS modules.
%The second background region was a polygon further to the north of the target,
% with low radio flux and few X-ray point sources.
With the background region thus selected,
 we excluded X-ray point sources from the region in the same way
 as was done for the source region.
Fig.~\ref{fig:0917+75} shows the resulting background annulus.
We then applied this local background region to the filtered images and
 exposure maps described in \S\ref{sec:filtering},
 and obtained the total background count,
 mean exposure, and background count rate scaled to the total exposure
 in the source region.
%To obtain an upper limit of source flux,
% we recalculate the background level with multiple background regions
% at the vicinity of the original, and obtain the standard deviation $\sigma$
% of this sample.
To assess the variation of the background level in our observation,
 we obtained additional background regions in the vicinity of the annulus
 and scaled the event rates within them in a similar fashion,
 and calculated the standard deviation $\sigma$ of scaled event rates
 in this sample.

To assess the amount of degradation in our data due to residual flares,
 we calculated the count rates of individual pointings that make up
 the blank sky data set of \citet{Nevalainen}.
 We then compared the background variation (over space) in our data
 with the variation of the blank-sky count rates (over time).
We found the variation in our background to be greater
 than that in the blank sky data by factors of 1.26, 1.27 and 2.05
 for the MOS1, MOS2 and and PN detectors, respectively.
%The flux limit we set in this study is worsened accordingly.
The flux limit we set in this study is affected accordingly.

\subsubsection{1401$-$33}
%^^^^^^^^^^^^^^^^^^^^^^^^
Compared to 0917+75, the nature of the background
 in our observation of 1401$-$33 is quite different.
The observation was made at a time of average solar wind activity,
as shown by \emph{ACE} data during this period (see Fig.~\ref{fig:ace});
 the overall count rates above about 10\,keV are
 also well within the accepted quiescent levels.
However, an X-ray halo of thermal emission centred at the neighbouring
 bright galaxy NGC~5419 about 10~arcmin away west of northwest
 \citep{Subrahmanyan2003}
 dominates the local background, producing excess flux across the entire
 field of view when compared to blank sky data.
This thermal emission is so strong that spectral lines caused by the solar wind,
 which are normally present below 1.0\,keV, are not observable.
%Given this dominating thermal component across the field of view,
% comparison with a standard background data set is meaningless.
%Instead, we made the assumption that the thermal emission is spherically
To assess the amount of thermal emission from the halo
 in our source region for 1401$-$33,
 we made the assumption that the thermal emission is spherically
 symmetric about the centre of the cluster (ie, NGC~5419),
 and chose a local background region equidistant from the centre
 of the thermal emission as the source region. % on the focal plane.
Our assumption is based on a previous X-ray observation of the cluster
 with the \emph{ROSAT} PSPC (in which IC/CMB emission from the relic was
 not detected); fig.~8 in \citet{Subrahmanyan2003} is an X-ray image
 from the observation, showing rotationally symmetric contours about NGC~5419.
% with about 8\% of the peak emission in both our source and background regions.
%\footnote{Do I need to give a number here, eg, the brightness level?}
%We acknowledge that such a background region is far from ideal,
% as our results depend heavily on the validity of the symmetry assumption.
%However, given the dominance of the thermal emission, this appears to be
% a reasonable assumption to first order.
The regular symmetric shape of the cluster justifies our choice
 of the background region.
%In addition, because a large portion of the X-ray field of view overlaps
% with the extent of the radio source at 330\,MHz \citep{Subrahmanyan2003},
% we chose a circular region just to the north of the target,
% instead of the usual annulus surrounding the source region.
% (but see \S\ref{sec:results} for a caveat).
In addition, \citet{Subrahmanyan2003} also revealed that
 a large portion of the \emph{XMM-Newton} field of view in our observation
 overlaps with the extent of the radio source at 330\,MHz.
Thus, we chose a circular region just to the north of the target,
 instead of the usual annulus surrounding the source region.
Fig.~\ref{fig:1401-33} shows the resulting background region,
 with the extent of the 330-MHz emission for reference.
To calculate the local background event rate and its variation,
 we applied the background region to the filtered image and exposure map,
 and obtained additional background regions, following the same recipe
 as outlined above for 0917+75.

%Another issue with our background region for 1401$-$33 is vignetting.
Because our background region for 1401$-$33 is not close to the centre
 of the telescopes' field of view, vignetting is a potential issue.
\citet{PradasKerp2005} reported an overcorrection of vignetting towards
 the rim of each detector by the routine \emph{eexpmap} that generates
 exposure maps in SAS versions 5.3.3 and 5.4.1.
%\footnote{FYI, I emailed Pradas, but got an `unknown recipient' mail error.
% I then emailed Kerp, but did not get a reply.}
To avoid overlapping with the extent of the radio emission of 1401$-$33
 at 330\,MHz, we have no choice but to place our background region
 at the periphery of the \emph{XMM-Newton} field of view.
Thus, the overcorrection of vignetting affects our analysis,
 potentially overestimating the count rate in the background region.
%We further discuss this issue at the end of the next section.
%\footnote{I could compare the counts in the peripheral polygon
% and annulus regions of 0917+75, assume those are equal, and infer
% the amount of overcorrection, but I have not done that yet, and so,
% I have no number to put in here.  However, the change will be relatively
% simple.  Probably just adding an extra sentence.  I have emailed J. Kerp
% in the meantime.}
On the other hand, the particle background that dominates at high energies
 is not vignetted.
To correct for vignetting of the local background properly,
 we followed a two-step background subtraction method
 described in Appendix~A of \citet{Arnaud2002}:
We calculated event rates in the blank-sky data from \citet{Nevalainen}
 at pixels in the source and background regions
 and within the optimal energies in this study.
We scaled these source- and background-region event rates
 according to the ratio of hard-band rates in our data
 and in the blank-sky data.
We then subtracted the scaled rates from the event rates in our data
 to remove the particle background from both the source and background regions.
Having eliminated the non-vignetted component,
 we then scaled the event rate in the background region to the source region
 according to the ratio of summed exposure in each region,
 for vignetting correction.
Finally, we arrived at the source event rate
 by substracting the scaled background-region rate from the source-region rate.
Table~\ref{tab:resultsCalculated} shows the relevant numbers in this analysis.

\subsubsection{Flux and field limits}
%^^^^^^^^^^^^^^^^^^^^^^^^^^^^^^^^^^^^
\label{sec:results}
Table~\ref{tab:resultsCalculated} summarizes our quantitative results.
%In the following paragraphs, we discuss only results from the MOS detectors,
In this paper, we discuss only results from the MOS detectors,
 as the MOS and PN results yield the same conclusions, and the MOS results
 provide more stringent limits.
%\placetable{tab:resultsCalculated}
%From these numbers, we see that the flux obtained from the two source regions
% is consistently lower than those from the background region
% in each observation, implying non-detection for all three observations.
% are both substantially lower than the local background levels;
%\footnote{Fiona, the source flux levels are 16--36$\sigma$
% below the scaled background levels.  So, I cannot simply say
% that `the source flux levels are consistent with the background levels
% to within N$\sigma$,' as you suggested.}
For 0917+75,
% the flux in the source region is consistent
% with that in the background region to within 3$\sigma$;
% the area-scaled event rates in the source and background regions are
% both 0.018~counts/sec (0.2--3.7\,keV), agreeing to better than 10\%.
 the event rates within the source region in the optimal band of 0.2--3.7\,keV
 are 0.0186 and 0.0176~counts~s$^{-1}$ for MOS1 and MOS2, respectively,
 while the corresponding area-scaled event rates in the background regions are
 $(0.0180 \pm 0.0016)$ and $(0.0194 \pm 0.0015)$~counts~s$^{-1}$, respectively.
%For 1401$-$33, the event rates in the on-axis source region
% (0.0347 and 0.0370~counts~s$^{-1}$, respectively) are
% only about 70 per cent of the area-scaled rates in the off-axis background region
% [$(0.0556 \pm 0.0041)$ and $(0.0568 \pm 0.0035)$~counts~s$^{-1}$, respectively].
For 1401$-$33, the blank-sky substracted event rates
 in the on-axis source region
 (0.0142 and 0.0168~counts~s$^{-1}$, respectively) are
 only a fraction of the blank-sky subtracted and area-scaled rates
 in the off-axis background region
 [$(0.0310 \pm 0.0043)$ and $(0.0319 \pm 0.0035)$~counts~s$^{-1}$, respectively].
%In other words, we did not detect the IC/CMB emission from either source.
These numbers, together with the appearance of the X-ray images
 in Figs.~\ref{fig:0917+75} and~\ref{fig:1401-33}, indicate that
% there is clearly not a detection of the IC/CMB emission from either source.
 there is not a detection of the IC/CMB emission from either source.

%We used PIMMS to convert the photon count rate to energy flux,
% and taking into account galactic absorption in this process.
% we converted the count rate to energy flux with PIMMS \citep{pimms},
% taking redshift and galactic absorption into account in the process.
We converted the measured event rate to the unabsorbed energy flux
 with PIMMS \citep{pimms}, assuming a power-law spectrum and
 accounting for redshift and galactic absorption.
We set the X-ray spectral index to be the same
 as the radio spectral index~\citep{1979MNRAS.188...25H},
 assuming any X-ray emission to be IC/CMB in origin.
%The assumed galactic \ion{H}{1} column densities were
%The assumed galactic \mbox{H\,{\sc i}} column densities were
We took the spectral indices ($\alpha=1.0$ for 0917+75 and 1.4 for 1401$-$33)
 and redshifts ($z=0.125$ and 0.0136, respectively)
 from previous radio measurements,
 and the galactic \mbox{H\,{\sc i}} column densities
 ($2.09 \times 10^{20}$ and $5.44 \times 10^{20}\,cm^{-2}$)
 from the FTOOLS programme nH.
% Table~\ref{tab:source param} gives their values.
%Note that PIMMS is designed for the study of point sources only,
%and it expects the input count rate to be that
%from a point source.\cite{PIMMSmanual}
%Thus, we scale the count rates measured from the extended sources
% by the ratio of the XMM point spread function (15'') to the size
% of each source, before entering the numbers in PIMMS.

%To obtain an upper limit of the source flux,
% we calculated the background variation in each observation
% with multiple background regions
% at the vicinity of the original, and obtained the standard deviation $\sigma$
% of this sample.
\iffalse
To assess the maximum source flux,
 we adopted twice the r.m.s. variation in the local background as an upper limit,
 to make direct comparison with previous studies possible.
When converted to energy flux at 0.3--10.0\,keV, the +2$\sigma$ upper limits
 are $3.48 \times 10^{-14}$\,erg~s$^{-1}$~cm$^{-2}$ for 0917+75
 and $9.71 \times 10^{-14}$\,erg~s$^{-1}$~cm$^{-2}$ for 1401$-$33.
The measured flux in the source region of 0917+75 is consistent with this
 2$\sigma$ limit, while the source flux for 1401$-$33 is as much as 5$\sigma$
 below the measured mean background.
\fi
To assess the maximum source flux, we adopted
 three times the r.m.s. variation in the local background as an upper limit.
When converted to energy flux at 0.3--10.0\,keV, the +3$\sigma$ upper limits
 are $5.22 \times 10^{-14}$\,erg~s$^{-1}$~cm$^{-2}$ for 0917+75
 and $1.47 \times 10^{-13}$\,erg~s$^{-1}$~cm$^{-2}$ for 1401$-$33.
The measured flux in the source region of 0917+75 is consistent with this
 3$\sigma$ limit, while the source flux for 1401$-$33 is about 4$\sigma$
 below the measured mean background.
Using Equation~11 in \citet{1979MNRAS.188...25H},
% Equation~9 in \citet{1998PASJ...50..389H},
 and radio measurements reported in \citet{HarrisSternWillis1993}
 and \citet{GossMcadamWellington1987},
% we obtained 2$\sigma$ lower limits on the magnetic field strength
 we obtained 3$\sigma$ lower limits on the magnetic field strength
 at each source;
% they are 0.93\,$\mu$G for 0917+75 and 1.9\,$\mu$G for 1401$-$33
 they are 0.81\,$\mu$G for 0917+75 and 2.2\,$\mu$G for 1401$-$33.
% (see Table~\ref{tab:resultsInferred} for further details).
%\input{tab3}
For comparison, the field strength obtained with the `classical' equipartition
 formula (with a low-frequency cutoff of the synchrotron emission
 at $\nu=10\,\mathrm{MHz}$)
% are 0.5\,$\mu$G and 0.8\,$\mu$G for 0917+75 and 1401$-$33, respectively.
 are 0.63\,$\mu$G and 1.3\,$\mu$G for 0917+75 and 1401$-$33, respectively.
Table~\ref{tab:resultsInferred} provides further details on these calculations.
\begin{table*}
\begin{minipage}{171mm} %\begin{center}
\renewcommand{\thefootnote}{\thempfootnote}

\caption{Physical parameters inferred from observations.
\label{tab:resultsInferred}}

%\tiny
%\scriptsize
%\small
\begin{tabular}{rr@{}l@{}lr@{}l@{}lr@{}l@{}lr@{}l@{}ll}
%\tableline
\hline %\tableline
%Target	& \multicolumn{2}{c}{0917+75\tablenotemark{a}}
%	& \multicolumn{2}{c}{1401$-$33\tablenotemark{b}} \\
Target	& \multicolumn{6}{c}{0917+75\footnote{\citealt{HarrisSternWillis1993}}}
	& \multicolumn{6}{c}{1401$-$33\footnote
					{\citealt{GossMcadamWellington1987}}} \\
Measurement
%	& MOS & PN			& MOS & PN \\
	& \multicolumn{3}{c}{MOS} & \multicolumn{3}{c}{PN}
	& \multicolumn{3}{c}{MOS} & \multicolumn{3}{c}{PN} \\
\hline %\tableline

Equipartition: $B_\mathrm{eq} \mapsto \mathrm{flux}$  \hfill \hspace{0pt} \\

Energy spectral index, $\alpha$ ($S_\nu \propto \nu^{-\alpha}$)
%	& \multicolumn{2}{c}{1.0}	& \multicolumn{2}{c}{1.4} \\
	& \multicolumn{3}{r@{}}{1}&\multicolumn{3}{@{}l}{.0}
	& \multicolumn{3}{r@{}}{1}&\multicolumn{3}{@{}l}{.4} \\
$c_{12}(\alpha)$\footnote{$c_1$, $c_5(2\alpha+1)$ and $c_{12}(\alpha)$ are defined by \citet[][Appendix~2, pp.~231--233]{Pacholczyk1970}.}
%	& \multicolumn{2}{c}{$9.3 \times 10^7$}
%	& \multicolumn{2}{c}{$1.6 \times 10^8$} \\
	& \multicolumn{3}{r@{}}{9}&\multicolumn{3}{@{}l}{.3 $\times 10^7$}
	& \multicolumn{3}{r@{}}{1}&\multicolumn{3}{@{}l}{.6 $\times 10^8$} \\
%Radio luminosity, $\log_{10}(L_r)$\tablenotemark{c}
Radio luminosity, $\log_{10}(L_r)$\footnote
		{Values were from table~4 of \citealt{HarrisSternWillis1993},
		 scaled to the currently accepted cosmology.
		 They assumed $H_0 = 50$~km~s$^{-1}$Mpc$^{-1}$, $q_0 = +0.5$,
		  $\Omega_\Lambda = 0$,
		  a lower cutoff frequency of 10~MHz, and simple spherical
		  and cylindrical geometries for the volumes of emission.
		 The original values were
		  $\log_{10}(L_r)=41.6$,
		  $\log_{10}(V)=72.8$ and
		  $B_\mathrm{eq}=0.5$\,$\mu$G for 0917+75, and
		  $\log_{10}(L_r)=41.3$,
		  $\log_{10}(V)=71.7$ and
		  $B_\mathrm{eq}=0.8$\,$\mu$G for 1401$-$33,
		  although a recalculation gave
		  $B_\mathrm{eq}=1.1$\,$\mu$G for 1401$-$33 instead.}
%	& \multicolumn{2}{c}{41.4}	& \multicolumn{2}{c}{41.0}
	& \multicolumn{3}{r@{}}{41}&\multicolumn{3}{@{}l}{.4}
	& \multicolumn{3}{r@{}}{41}&\multicolumn{3}{@{}l}{.0}
							& erg~s$^{-1}$ \\
Emitting volume, $\log_{10}(V)$\footnotemark[\value{mpfootnote}]
%	& \multicolumn{2}{c}{72.4}	& \multicolumn{2}{c}{71.3}
	& \multicolumn{3}{r@{}}{72}&\multicolumn{3}{@{}l}{.4}
	& \multicolumn{3}{r@{}}{71}&\multicolumn{3}{@{}l}{.3}
							& cm$^3$ \\
Magnetic field strength, $B_\mathrm{eq}$\footnote
		{Values were calculated using
		the classical equipartition formula:
$B_\mathrm{eq} = \left[ \frac{8 \pi (1+K) c_{12} L_r}{\phi V} \right]^{2/7}$.}
%	& \multicolumn{2}{c}{0.63}	& \multicolumn{2}{c}{1.3} & $\mu$G \\
	& \multicolumn{3}{r@{}}{0}&\multicolumn{3}{@{}l}{.63}
	& \multicolumn{3}{r@{}}{1}&\multicolumn{3}{@{}l}{.3} & $\mu$G \\
%IC/CMB flux (0.3--10\,keV) comparison: \hfill \hspace{0pt} \\
%Flux comparison (0.3--10\,keV): \hfill \hspace{0pt} \\
%at equipartition\tablenotemark{d}
IC/CMB flux (0.3--10\,keV)\footnote{Values were calculated using
			Equation~11 in \citealt{1979MNRAS.188...25H}:
$B^{\alpha+1} = \frac{	(5.05 \times 10^4)^\alpha
			C(\alpha) G(\alpha) (1+z)^{\alpha+3} S_r \nu_r^\alpha}
		     {	10^{47} S_x \nu_x^\alpha}$,
			with $G(\alpha)$ from table~1 therein.}
%  Our \emph{XMM-Newton} proposal\tablenotemark{d}
%	& \multicolumn{2}{c}{$1.8 \times 10^{-13}$}
%	& \multicolumn{2}{c}{$3.1 \times 10^{-12}$} & erg~s$^{-1}$cm$^{-2}$ \\
%  \citealt{1979MNRAS.188...25H}
%	& \multicolumn{2}{c}{$1.2 \times 10^{-13}$}
%	& \multicolumn{2}{c}{$1.4 \times 10^{-12}$} & erg~s$^{-1}$cm$^{-2}$ \\
%	& \multicolumn{2}{c}{$5.4 \times 10^{-13}$} & erg~s$^{-1}$cm$^{-2}$ \\
	& \multicolumn{3}{r@{}}{8}&\multicolumn{3}{@{}l}{.7 $\times 10^{-14}$}
	& \multicolumn{3}{r@{}}{5}&\multicolumn{3}{@{}l}{.5 $\times 10^{-13}$}
							& erg~s$^{-1}$cm$^{-2}$ \\
%  \citealt{1998PASJ...50..389H}
%	& \multicolumn{2}{c}{$2.9 \times 10^{-13}$}
%	& \multicolumn{2}{c}{$3.3 \times 10^{-12}$} & erg~s$^{-1}$cm$^{-2}$ \\
This study:
	flux $\mapsto B_\mathrm{measured}$ \hfill \hspace{0pt} \\
Energy spectral index, $\alpha$ ($S_\nu \propto \nu^{-\alpha}$)
%	& \multicolumn{2}{c}{1.0}	& \multicolumn{2}{c}{1.4} \\
	& \multicolumn{3}{r@{}}{1}&\multicolumn{3}{@{}l}{.0}
	& \multicolumn{3}{r@{}}{1}&\multicolumn{3}{@{}l}{.4} \\
\setcounter{mpfootnote}{3}
$C(\alpha) = \sqrt{c_1} / \left[ 4 \pi 2^\alpha c_5(2\alpha+1) \right]$\footnotemark[\value{mpfootnote}]
%	& \multicolumn{2}{c}{$1.32 \times 10^{31}$}
%	& \multicolumn{2}{c}{$1.30 \times 10^{31}$} \\
	& \multicolumn{3}{r@{}}{1}&\multicolumn{3}{@{}l}{.32 $\times 10^{31}$}
	& \multicolumn{3}{r@{}}{1}&\multicolumn{3}{@{}l}{.30 $\times 10^{31}$} \\
\setcounter{mpfootnote}{6}
$G(\alpha)$\footnotemark[\value{mpfootnote}]
%	& \multicolumn{2}{c}{0.500}	& \multicolumn{2}{c}{0.653} \\
	& \multicolumn{3}{r@{}}{0}&\multicolumn{3}{@{}l}{.500}
	& \multicolumn{3}{r@{}}{0}&\multicolumn{3}{@{}l}{.653} \\
Redshift, $z$
%	& \multicolumn{2}{c}{0.125}	& \multicolumn{2}{c}{0.0136} \\
	& \multicolumn{3}{r@{}}{0}&\multicolumn{3}{@{}l}{.125 }
	& \multicolumn{3}{r@{}}{0}&\multicolumn{3}{@{}l}{.0136} \\
%Synchrotron flux density at 1~Hz
%	& \multicolumn{2}{c}{$1.8 \times 10^{-15}$}
%	& \multicolumn{2}{c}{$3.2 \times 10^{-11}$}
%					& erg~s$^{-1}$cm$^{-2}$Hz$^{-1}$ \\
Synchrotron flux density, $S_r(\nu_r)$
%	& \multicolumn{2}{c}{1.2}	& \multicolumn{2}{c}{0.5} & Jy \\
	& \multicolumn{3}{r@{}}{1}&\multicolumn{3}{@{}l}{.2}
	& \multicolumn{3}{r@{}}{0}&\multicolumn{3}{@{}l}{.5} & Jy \\
Frequency of radio measurement $\nu_r$
%	& \multicolumn{2}{c}{0.151}	& \multicolumn{2}{c}{1.49} & GHz \\
	& \multicolumn{3}{r@{}}{0}&\multicolumn{3}{@{}l}{.151}
	& \multicolumn{3}{r@{}}{1}&\multicolumn{3}{@{}l}{.49 } & GHz \\
%IC/CMB flux (0.3--10\,keV), 3-$\sigma$ upper limit
3-$\sigma$ IC/CMB flux (0.3--10\,keV)
%measured 2$\sigma$ upper limit
%	& $\leq 3.48 \times 10^{-14}$
%	& $\leq 1.27 \times 10^{-13}$
%					& $\leq 9.71 \times 10^{-14}$
%					& $\leq 1.19 \times 10^{-13}$
%						& erg~s$^{-1}$cm$^{-2}$ \\
%measured 3$\sigma$ upper limit
	& $\leq$ 5&.22 & $\times 10^{-14}$
	& $\leq$ 1&.91 & $\times 10^{-13}$
					& $\leq$ 1&.47 & $\times 10^{-13}$
					& $\leq$ 1&.49 & $\times 10^{-13}$
						& erg~s$^{-1}$cm$^{-2}$ \\
%X-ray flux normalization, $S_x \nu_x^\alpha = \mathrm{flux} / \int \nu^{-\alpha}\,d\nu$
$S_x \nu_x^\alpha = \mathrm{flux} / \int \nu^{-\alpha}\,d\nu$
	& 1&.5 & $\times 10^{-14}$ & 5&.4 & $\times 10^{-14}$
	& 4&.3 & $\times 10^{-7}$  & 4&.4 & $\times 10^{-7}$ \\
%magnetic field strength, 3$\sigma$ lower limit,
3$\sigma$ magnetic field strength,
 $B_\mathrm{measured}$\footnotemark[\value{mpfootnote}]
%	& $\geq 0.81$  & $\geq 0.42$	& $\geq 2.2$  & $\geq 2.2$ & $\mu$G \\
	& $\geq$ 0&.81 & & $\geq$ 0&.42 &	& $\geq$ 2&.2 & & $\geq$ 2&.2 & & $\mu$G \\

\iffalse
Magnetic field strength: \hfill\ \hspace{0pt} \\
%at equipartition, $B_{eq}$\tablenotemark{c}
\setcounter{mpfootnote}{3}
at equipartition, $B_{eq}$\footnotemark[\value{mpfootnote}]
	& \multicolumn{2}{c}{0.5}	& \multicolumn{2}{c}{0.8} & $\mu$G \\
%Magnetic field strength, $B_{\mathit measured}$, 3$\sigma$ lower limit
%measured 2$\sigma$ lower limit, $B_{\mathit measured}$
%%  \citealt{1979MNRAS.188...25H}
%	& $\geq 0.93$  & $\geq 0.49$	& $\geq 1.9$  & $\geq 1.7$ & $\mu$G \\
%%  \citealt{1998PASJ...50..389H}
%%	& $\geq 1.2$  & $\geq 0.62$	& $\geq 2.4$  & $\geq 2.2$ & $\mu$G \\
measured 3$\sigma$ lower limit, $B_{\mathit measured}$
%  \citealt{1979MNRAS.188...25H}
	& $\geq 0.75$  & $\geq 0.39$	& $\geq 2.1$  & $\geq 2.0$ & $\mu$G \\
\fi

\iffalse
$B_{\mathit measured} / B_{eq}$
%  \citealt{1979MNRAS.188...25H}
	& $\geq 1.5$  & $\geq 0.79$	& $\geq 1.9$  & $\geq 1.8$  \\
%  \citealt{1998PASJ...50..389H}
%	& $\geq 2.4$  & $\geq 1.2$	& $\geq 3.0$  & $\geq 2.7$  \\
$(1+K) / \phi V \propto B^{7/2}$ (See text, \S\ref{sec:discussion})
%  \citealt{1979MNRAS.188...25H}
	& $\geq 4.3$   & $\geq 0.4$	& $\geq 10$ & $\geq 7.1$ \\
%  \citealt{1998PASJ...50..389H}
%	& $\geq 20$   & $\geq 2.1$	& $\geq 45$ & $\geq 32$ \\
\fi
\hline %\tableline
\end{tabular}

\end{minipage} %\end{center}
\end{table*}

Alternatively, using a revised equipartition formula
 from \citet*[Equation~A4]{BrunettiSettiComastri1997},
% with the low-energy electron population cutoff at $\gamma = 100$,
% the field strengths are 0.6\,$\mu$G and 1.0\,$\mu$G, respectively.
 with the low-energy electron population cutoff
 at $\gamma_\mathrm{min} = 1069 \sqrt{E_\mathrm{x,min}/\mathrm{keV}} = 500$
 \citep{1979MNRAS.188...25H} to match the energy range of \emph{XMM-Newton},
 and with $D(\delta) \sim 1$,
 the field strengths are 0.79\,$\mu$G and 1.6\,$\mu$G, respectively.
We note, however, that these values are very sensitive
 to the choice of $\gamma_\mathrm{min}$,
 whose true value is unknown without better knowledge
 of the low-energy synchrotron spectrum.

\section{DISCUSSION}
%-------------------
\label{sec:discussion}
%In this section, we first give some background information on each target,
% and then compare our flux limits with any previous measurements
% and with the equipartition estimates.
In this section, we first discuss each target separately,
 and then comment on the flux and field limits.

%\subsection{Source by source considerations}
%%-------------------------------------------
\subsection{0917+75}
\label{sec:0917+75}
\citet{DewdneyCostainMchardy1991} originally thought
 that 0917+75 was the radio halo of an uncatalogued cluster of galaxies,
 but \citet{HarrisSternWillis1993} recategorized it
% as a relic radio galaxy on the basis of its high fractional polarization.
 as a relic radio galaxy on the basis of its high polarization.
It is apparently associated with two galaxies of redshift $z = 0.125$
%, in a poor cluster
 within the Rood~27 supercluster,
 whose members include the Abell clusters A762, A786, and A787.
\citet{HarrisSternWillis1993} reported
 a spectral flux density of $S_r(\nu_r) = (1.2 \pm 0.2)\,\mathrm{Jy}$
 at $\nu_r=151\,\mathrm{MHz}$,
 and an energy spectral index of $\alpha = 1.0$ at low frequencies,
% with a cutoff or spectral break close to 150\,MHz.
 with a spectral break close to 150\,MHz.
The spectrum flattens to $\alpha = 0.6$ between 325 and 1500\,MHz,
 and steepens again above that.
This complex spectral shape could be the superposition of synchrotron emission
 from two populations of relativistic electrons;
 it would be the lower-energy population
 ($478 < \gamma < 3380$, $1\,\mathrm{MHz} < \nu_r < 48\,\mathrm{MHz}$)
 that is responsible for IC/CMB emission
 in the energy range of \emph{XMM-Newton} (0.2--10.0\,keV).
Unfortunately, the spatial distribution of the lower-energy population is
 as yet undetermined (below 300\,MHz), due to limited angular resolution
 in low-frequency measurements in the past.
In this study, we have made the assumption that the IC/CMB emitting electrons
 have the same spatial extent as those in the higher-energy population,
 which was given by the NVSS radio maps at 1.4\,GHz.
\iffalse
\footnote{Can we improve on this?  eg, is the spatial extent given by the
 325\,MHz map from WSRT the same as the one from NVSS at 1.4 GHz?
 Are there newer measurements at lower frequencies available now?
 The 151\,MHz measurement was inferred from the 6C survey plus some
 higher-resolution data that excludes the neighbouring source R5,
 but the relic was not resolved?
 There were also measurements made w/ TPT at 30.9 and 57.4\,MHz,
 but the beam size was comparable or larger than the relic.
 Can we improve on any of these?  Note that the calculations in this paper
 are based on 1.2\,Jy at 151\,MHz and $B_{eq}$ reported
 in \citet{HarrisSternWillis1993} only
 (and the $+5\sigma$ contour in the NVSS map).}
\fi

\citet{HarrisWillisDewdney1995} observed 0917+75
 in a 15.8-ks \emph{ROSAT} PSPC observation
 (with the same assumption on the spatial extent of the IC/CMB electrons).
They found the source isolated in X-ray,
 and not contaminated by thermal emission.
They placed a $2\sigma$ upper limit
 of $2.3 \times 10^{-14}$\,erg~s$^{-1}$cm$^{-2}$ (27~counts / 15\,827\,s)
%$1.7 \times 10^{-3}$~photons~s$^{-1}$
 on the 0.5--2.0\,keV emission from the region.
%Our result in this study improved on this limit:
With the better sensitivity of \emph{XMM-Newton},
our result improves on this limit:
% When converted to 0.5--2.0\,keV, our $2\sigma$ upper limit becomes
 when converted to 0.5--2.0\,keV and $2\sigma$, our upper limit becomes
 $1.38 \times 10^{-14}$\,erg~s$^{-1}$cm$^{-2}$.
The $2\sigma$ minimum magnetic field strength also
% increases from 0.5\,$\mu$G (the equipartition value) to 0.76\,$\mu$G,
% increases from 0.76\,$\mu$G at 2$\sigma$ (=0.60\,$\mu$G at 3$\sigma$)
% to 0.76\,$\mu$G at 3$\sigma$,
% increases from 0.76\,$\mu$G to 0.93\,$\mu$G,
 increases from 0.76\,$\mu$G to 0.99\,$\mu$G,
 as inferred from this and the radio measurements,
 but the difference is small compared to uncertainties
 in the equipartition value of 0.63\,$\mu$G.
% of 0.76\,$\mu$G inferred from this and the radio measurements above exceeded
% the nominal equipartition value by a small factor of 1.5.
\iffalse
We note that the background level in our observation was
 higher than usual, due to residual flares
 and high solar wind activity (see \S\ref{sec:analysis}).
Thus, one can very likely improve the upper flux limit obtained in this study
 with additional \emph{XMM-Newton} observations.
%\footnote{Regarding the similarity of our limit and that in Harris et al.(1995),
% I don't really have a convincing explanation.  Have pointed out in the Results
% section that flares worsened the flux limit by about 25\%, but with the two
% MOS detectors coadded, it reduces to 10\%.
% The effect on the magnetic field limit would be even smaller.
%Given that variations in the Nevalainen et al. data set give a similar limit,
% I still think that the lack of improvement is due to the extended source
% not getting the benefit of a much smaller HPD.
%Please let me know if you think I need to comment on this point here;
% otherwise, I shall leave it as is.}
\fi

\subsection{1401$-$33}
1401$-$33 extends over a $20~\mathrm{arcmin} \times 9~\mathrm{arcmin}$ region in the poor cluster
 Abell~S753 around NGC~5419.
The morphology is strongly suggestive of a relic radio galaxy.
However, the parent galaxy has not been identified.
\citet{GossMcadamWellington1987} reported
 a spectral flux density of $S_r(\nu_r) = 0.5\,\mathrm{Jy}$
 at $\nu_r = 1.4\,\mathrm{GHz}$,
 with a spectral index of $\alpha = 1.4$.
Alternatively,
 \citet{Subrahmanyan2003} reported $\alpha=1.4$ between 330\,MHz
 and 1.4\,GHz, possibly increasing to $\alpha=1.9$ between 1.4 and 2.4\,GHz,
 but only for the brightest part of the relic
 (the `NE rim', which we consider in this study).
When a large area of faint emission to the southwest was included,
 they reported $\alpha = 2.0$
 between 330\,MHz and 1.4\,GHz, and $\alpha = 2.9$ between 1.4 and 2.4\,GHz.
To reconcile with an 85-MHz measurement of 57\,Jy from 1960,
 they suggested that there is a spectral break between 100 and 300\,MHz.
Assuming a break at 165\,MHz, $\alpha = 0.7$ below the break frequency.
%\footnote{I actually do not understand this, because the data from the two
%papers agreed with each other.  Then, why would they come to different
%conclusions about $\alpha$?}
\iffalse
\footnote{\citet{Subrahmanyan2003} said that they were
 going to propose new VLA observation of a source close to 1401$-$33 at 74\,MHz.
 What is the current state of the art for radio measurements below 100\,MHz?
 (eg, angular resolution.)
 Can we observe 0917+75 again?  Perhaps for a future paper?}
\fi

The only \emph{ROSAT} observations of the region around 1401$-$33 were offset by 45~arcmin
 and are both less than 5\,ks long, placing no interesting limit
 on the magnetic field strength.
In our observation,
 we found an elevated flux level in the source region
 when compared to blank sky data,
 but no excess when compared to a local background region
 equidistant from the bright galaxy NGC~5419.
%\footnote{not exactly true as it is, but I can easily fix this...after writing this draft, which is taking forever.}
%\footnote{Actually, I am not exactly sure about this point.
%Because I do not know how to run XSPEC, and because we have assumed up to now
% that there are not enough X-ray counts to perform a proper spectral fit,
% we have not done so.  Maybe we should?  Yet, I don't want to spend time
% on this and get scooped in the meantime.}
\citet{Subrahmanyan2003} attributed this excess flux to thermal emission
 from a halo centred at NGC~5419.
When we inspect our data with XSPEC, we are able to confirm that this
 excess emission fits much better to a thermal spectrum than to a power-law
 spectrum, although the statistics are too low to produce a high-quality fit.
%However, we inspected the spectra of the neighbouring bright galaxy NGC~5419
% and of a blank sky region roughly equidistant from the galaxy as our target;
% we found that the elevated level is attributable to contamination
% by the bright source, and that there is no excess flux in the source region
% compared to its immediate vicinity.
%This is the strangest of the three non-detections,
% because the inverse Compton flux is the strongest amongst our three targets
% for a magnetic field at equipartition,
% and a non-detection implies a magnetic field strength
% significantly higher than the equipartition value.
%The result from this study places a $2\sigma$ upper limit on any excess
% X-ray flux specific to 1401$-$33
% at $9.71 \times 10^{-14}$\,erg~s$^{-1}$cm$^{-2}$ within 0.3--10.0~keV,
% and a minimum magnetic field of 1.9\,$\mu$G.
The result from this study places a $3\sigma$ upper limit on any excess
 X-ray flux specific to 1401$-$33
 at $1.47 \times 10^{-13}$\,erg~s$^{-1}$cm$^{-2}$ within 0.3--10.0~keV,
 and a minimum magnetic field of 2.2\,$\mu$G.
% beyond the old limit derived from the equipartition formula.
% larger than the old limit derived from the equipartition formula.
These are the first limits on IC/CMB emission reported for 1401$-$33.

\iffalse
\subsection{1253+275}
1253+275 is located 70' from the centre of the Coma radio halo,
 and is the largest of the three sources we observed (10'$\times$30').
The identity of the radio galaxy which produced the diffuse emission
 is uncertain; two possibilities are IC~3900 and NGC~4789.
Radio observations between 151 and 4750\,MHz can be fit by a single power law
 with energy spectral index $\alpha = 1.18$ and flux density normalization
 1.4\,Jy at 326\,MHz \citep{GiovanniniFerettiStanghellini1991}.
1253+275 has the least secure classification as a relic of our sample.

%Thermal contamination from thermal emission at the Coma outskirts
% is not a significant problem.

Our observation of 1253+275 was plagued by flares,
 evident from an inspection of the light curves of high-energy events.
The structure of this source, as seen in the radio data at 1.4\,GHz,
 is filamentary \citep{GiovanniniFerettiStanghellini1991}.
A filamentary mask is thus used accordingly, masking out small regions
 of low radio emission across the field.
(See Fig.?)
For the local background region, we picked the northeast corner
 of the field of view, as the radio emission at 608\,MHz extends towards
 the southeast and fuse with the emission from the neighbouring galaxy NGC~4789.
\fi

%\subsection{General digest}
%--------------------------
\subsection{On the flux and field limits}
%----------------------------------------
%\label{sec:discussion}

While the number of known relics has grown
 beyond 30~\citep*[see, e.g.,][for representative samples]{GiovanniniTordiFeretti1999,KempnerSarazin2001},
 studies of their IC/CMB emission remain scarce,
 and not a single convincing detection has been made so far.
% and not a single concrete detection has been made so far.
% despite IC/CMB being an obligatory process for all synchrotron sources.
Table~\ref{tab:reliclist} lists the relics whose IC/CMB emission has been studied,
 together with the reported field limits.
For consistent comparison,
 we have scaled the equipartition fields from previous studies
 to the currently acceptable cosmology,
 but we make no attempt to correct for differences in other parameters
 (eg, the frequency band integrated),
 as the required information is not always available.
\iffalse
; obj   H0,q0,Lambda    z       B_eq		H0_ic,q0_ic,Omega_Lambda
;A85     100,0.5,0      0.0555  1.3		50
;A754    50,0.5,0       0.0542  0.3		50
;A1367   100,0.5,0      0.022   2		50
;A2034   ---            0.113   ---		50
;A133    75,0.5,0       0.0562  14.4		70
;Coma    50,0.5,0       0.023   0.6		70,-0.55,0.7
;0917+75 50,0.5,0       0.125   0.5		70,-0.55,0.7
;1401-33 50,0.5,0       0.0136  0.8 x 1.1	70,-0.55,0.7
\fi

\begin{table*}
%\begin{minipage}{137mm} %\begin{center}
\begin{minipage}{170mm} %\begin{center}
\renewcommand{\thefootnote}{\thempfootnote}

\caption{List of relics with field limits from published IC/CMB measurements.
\label{tab:reliclist}}

%\tiny
%\scriptsize
%\small
\begin{tabular}{r@{}lr@{ }rclrr@{}lr@{}lcr@{}lcc}
%\tableline
\hline %\tableline
\multicolumn{2}{c}{Relic}	& \multicolumn{2}{c}{Host}
			& \multicolumn{7}{c}{Classical equipartition} &
			& \multicolumn{4}{c}{IC/CMB} \\
			\cline{5-11} \cline{13-16}
&				& \multicolumn{2}{c}{cluster}
			& Ref.\footnote
{References. -- (1)~\citealt{FerettiGiovannini1996};
 (2)~\citealt{1998MNRAS.296L..23B}; % {BagchiPislarLimaNeto1998};
 (3)~\citealt{BacchiFerettiGiovanniniGovoni2003};
 (4)~\citealt{Fusco-Femiano2003};
 (5)~\citealt{GavazziTrinchieri1983};
 (6)~\citealt{HenriksenMushotzky2001};
 (7)~\citealt*{KempnerSarazinMarkevitch2003};
 (8)~\citealt{Slee2001};
 (9)~\citealt{Fujita2004};
(10)~\citealt{HarrisSternWillis1993};
(11)~\citealt{FerettiNeumann2006};
(12)~this paper.}
			& \multicolumn{1}{c}{$z$}
			& $H_0$
			& \multicolumn{2}{c}{$B_\mathrm{eq}$ ($\mu$G)}
			& \multicolumn{2}{c}{$B_\mathrm{eq}$ ($\mu$G)} &
			& \multicolumn{2}{c}{$B_\mathrm{measured}$}
			& Instrument
			& Ref.\footnotemark[\value{mpfootnote}] \\
	& &		& & & &
			& \multicolumn{2}{c}{in references\footnote{All cited $B_\mathrm{eq}$ values assume
							$q_0 = +0.5$, $\Omega_\Lambda = 0$, and $H_0$ as shown above.}}
			& \multicolumn{2}{c}{cosmology updated} &
			& \multicolumn{2}{c}{($\mu$G)} \\
\hline %\tableline

~0038&$-$096			& ~A &   85~
	& \,\,1 & 0.0555 & 100 & 1&.3	& 1&.1(6) & & 0&.95\footnote
	{IC/CMB detection has since been refuted by later studies.}	& \emph{ROSAT}      & \,\,2 \\
\multicolumn{2}{l}{`East relic'}& ~A &  754~
	& \,\,3 & 0.0542 &  50 & 0&.3	& 0&.3(3) & & 0&.1\footnotemark[\value{mpfootnote}]
									& \emph{BeppoSAX}   & \,\,4 \\
~1140&+203			& ~A & 1367~
	& \,\,5 & 0.022  & 100 & 2&		& 1&.(8)	& & $>$0&.84		& \emph{RXTE}       & \,\,6 \\
\multicolumn{2}{l}{`A~2034'}	& ~A & 2034~
	&  $\cdots$ & 0.113  & $\cdots$ & \multicolumn{2}{c}{$\cdots$}	& \multicolumn{2}{c}{$\cdots$}
									& & \multicolumn{2}{c}{0.3--0.9\footnote
	{The authors commented that the errors are `admittedly quite large.'}}	& \emph{Chandra} & \,\,7 \\
\multicolumn{2}{l}{`A~133\_7a'}	& ~A &  133~
	& \,\,8 & 0.0562 &  75 & \,\,\,\,\,\,\,\,\, 14&.4\footnote
		{Minimum energy instead of equipartition is assumed.}
		& \,\,\,\,\,\,\,\,\,\,\,\,\,\,\,\,\, 14&.0 & & \,\, $>$1&.5	& \emph{XMM-Newton} & \,\,9 \\
~1253&+275			& \multicolumn{2}{c}{Coma}
	&    10 & 0.023  &  50 &  0&.6	& 0&.6(6) & & $>$1&.05		& \emph{XMM-Newton} &    11 \\
~0917&+75			& \multicolumn{2}{c}{Rood~27}
	&    10 & 0.125  &  50 &  0&.5	& 0&.63 & & $>$0&.81		& \emph{XMM-Newton} &    12 \\
~1401&$-$33			& ~A & S753~
	&    10 & 0.0136 &  50 &  1&.1\footnote
		{Recalculation of $B_\mathrm{eq}$ using luminosity and volume from Ref.(12)
		gives this value instead 0.8\,$\mu$G from Ref.(12).}	&  1&.2(5) & & $>$2&.2	& \emph{XMM-Newton} &    12 \\
\hline %\tableline
\end{tabular}

\iffalse
\tablenotetext{a}{IC/CMB detection has since been refuted by later studies.}
\tablenotetext{b}{The authors commented that the errors are `admittedly quite large.'}
\tablenotetext{c}{Minimum energy instead of equipartition is assumed.}
\tablenotetext{d}{References.---(1) \citealt{FerettiGiovannini1996};
 (2) \citealt{1998MNRAS.296L..23B}; % {BagchiPislarLimaNeto1998};
 (3) \citealt{BacchiFerettiGiovanniniGovoni2003};
 (4) \citealt{Fusco-Femiano2003};
 (5) \citealt{GavazziTrinchieri1983};
 (6) \citealt{HenriksenMushotzky2001};
 (7) \citealt*{KempnerSarazinMarkevitch2003};
 (8) \citealt{Slee2001};
 (9) \citealt{Fujita2004};
(10) \citealt{HarrisSternWillis1993};
(11) \citealt{FerettiNeumann2006};
(12) this paper.}
\fi

\end{minipage} %\end{center}
\end{table*}

%There are three reports of detections to date,
There are three claims of possible detections to date,
 in the Abell clusters A~85, A~754 and A~2034.
\citet{1998MNRAS.296L..23B} detect IC/CMB emission
 from 0038$-$096 near Abell~85, using the PSPC on \emph{ROSAT}.
However, the statistics are poor, and
 the observation suffers from thermal contamination.
More recently, \citet{DurretLimaNetoForman2005} determine from a more
 sensitive observation with \emph{XMM-Newton}
 that the X-ray and radio emissions do not coincide spatially,
% and that the X-ray spectrum does not fit better to a nonthermal model
% than to a thermal model.
 and that the X-ray spectrum fits equally well to a thermal model
 with and without an added power-law component.
\citet{Fusco-Femiano2003} find nonthermal emission above 45\,keV
 in the aggregate spectrum of A~754 from the PDS on \emph{BeppoSAX};
 they attribute it to the central halo and `east relic' in the cluster,
 but do not rule out contamination from the radio galaxy 26W~020
 in the PDS field of view.
Subsequently, \citet{HenriksenHudsonTittley2004} are able
 to account for this hard X-ray component with a power-law model of 26W~020.
Finally, \citet{KempnerSarazinMarkevitch2003} find weak evidence
 for IC/CMB emission from the relic in A~2034 with \emph{Chandra},
 but comment that the errors on the measurement are quite large.
With the uncertanties in these studies,
 the nature of IC/CMB emission from relics remains elusive.

%None of these limits is more than a factor of a few from the values derived from the `classical' equipartition formula, which is well-known to contain multiple parameters with large uncertainties.
While none of the IC/CMB field strengths and upper limits in Table~\ref{tab:reliclist} is more than a factor of a few from the values derived from the `classical' equipartition formula,
 it is interesting, nevertheless, to note that the three most recent measurements -- on 0917+75, 1253+275 and 1401$-$33 -- all give lower limits for the field strength which are slightly larger than the equipartition values.
It is as yet unclear whether these deviations are significant,
 as is evident in the comparison of our field limits with the equipartition
 values obtained from the revised fomula of \citet{BrunettiSettiComastri1997}.
Both methods of determining the field strength
 involve parameters with large uncertainties, as we explain below.

The equipartition magnetic field, $B_{eq}$, is
 given by numerous authors in the past,
 for instance, as in \citet{HarrisWillisDewdney1995}:
$$
B_{eq} = \left[ \frac{8 \pi (1 + K) c_{12} L_r}{ \phi V} \right]^{2/7},
$$
 where $L_r$ is the radio luminosity, $V$ the emitting volume,
 $K$ the ratio of proton energy to electron energy,
 $\phi$ the filling factor for the emitting plasma, and
 $c_{12}$ a weak function of the spectral index and frequency band.
We note that uncertainty in the cosmological distance scale
 contributes to uncertainty in $L/V$.
In addition, we have assumed the most conservative values for $K$
 ($=0$) and $\phi$ ($=1$), which results in high X-ray flux
 and low $B$ estimates at equipartition.
In order to reconcile with the limits imposed by our results,
 $(1+K)/\phi$ has to be 2.4 times larger for 0917+75
 and 6.8 times larger for 1401$-$33.

%On the other hand, there are also uncertainties in the determination
% of the magnetic field from IC/CMB measurements.
In calculating the IC/CMB flux, we have assumed that we can infer the nature
 of the IC/CMB-emitting electrons from past radio observations,
 while this may not indeed be the case.
For instance, although the spectral indices for IC/CMB and synchrotron
 emissions should be the same, electrons responsible for IC/CMB emission
% in the energy range of \emph{XMM-Newton} (0.2--10.0\,keV)
% have $478 < \gamma < 3380$,
 in the energy range of \emph{XMM-Newton} (0.3--10.0\,keV)
 have $586 < \gamma < 3381$,
% and they emit synchrotron radiation at frequencies between 1 and 50\,MHz,
 and in a 1-$\mu$G magnetic field,
 they emit synchrotron radiation at frequencies between 1 and 50\,MHz,
 where radio observations are either poor in spatial resolution,
 or entirely unavailable from Earth, due to reflection off the ionosphere.
Thus, we can only rely on extrapolations of the spectral indices
 at higher frequencies to infer the IC/CMB index.
Yet, as is evident from the discussion above, radio spectra of relic galaxies
 often exhibit complex shapes with multiple breaks.
For both 0917+75 and 1401$-$33, there are indications that a break may exist
 in their radio spectra between 100 and 300\,MHz, and it is not guaranteed
 that these are the only ones.
If the spectrum of a synchrotron source flattens at low frequencies,
 that would imply lower IC/CMB flux as well, leading to closer agreement
 of IC/CMB measurements and equipartiton.
%\footnote{Note to self: Calculate how small alpha has to be in order
% to be consistent with our upper flux limits.}

%Various authors have proposed revisions to the equipartition formula:
Various authors have proposed alternatives
 to the classical equipartition formula:
\citet{BrunettiSettiComastri1997} integrate over a fixed particle energy range
 instead of a fixed synchrotron frequency range;
\citet{BeckKrause2005} replace the energy density ratio
 of ions to electrons with their number density ratio,
 and give estimates of the number density ratio for various types of objects;
\citet{PfrommerEnsslin2004} consider minimum energy and equipartition criteria
 in the particular scenario where the synchrotron-emitting
 electrons are produced in inelastic collisions between cosmic-ray protons
 and ambient thermal gas in the cluster.
%Yet, there is not enough observational evidence to date to support or refute these ideas.
Yet, uncertainties in the source geometry and electron population
 can only be resolved with new measurements.

\iffalse
It is interesting, nevertheless, to note that the three most recent measurements -- on 0917+75, 1253+275 and 1401$-$33 -- all give lower field limits slightly larger than the equipartition values.
These results contradict IC/CMB measurements in radio lobes,
 where recent studies claim field strengths slightly below
 equipartition instead~\citep*[e.g.][]{TashiroMakishimaKaneda2000,LeahyGizani2001,Croston2005}.
Because various uncertainties in IC/CMB measurements,
 including thermal contamination and
 potential breaks in the low-energy electron spectrum,
 all tend to underestimate the magnetic field,
 our findings cast doubts on the results from radio lobes.
To make progress, more X-ray studies are needed to get a broader picture
 of nonthermal emission from relics, and more sensitive observations are needed
 to better constrain their energy budget.
\fi

%Further progress on the study of IC/CMB emission from relics is only possible
% if we can better determine these parameters.
One way to better determine the various uncertain paremeters is
 to make new measurements at as yet unexplored wavelengths.
A new generation of low-frequency radio telescope arrays
 could provide radio measurements of relic galaxies
 at low frquencies (10--250\,MHz) in the future,
 thus better constraining both the spectral index
 and spatial extent of IC/CMB emitting particles.
Alternatively, a new hard X-ray / soft gamma-ray telescope would enable us
 to observe the IC/CMB emission by particles whose synchrotron emission is
 currently observable (particles emitting synchrotron radiation at 330\,MHz
 will emit IC/CMB radiation at 70\,keV),
 thus eliminating the problems of uncertain geometry
 and an uncertain spectral index.
Hard X-ray observations also have the virtue
 that thermal emission no longer dominates in this energy range,
 so that sources thermally contaminated in soft X-rays can be observed,
 substantially increasing the sample size.
Better determination of the filling factor $\phi$ could come
 from radio measurements of relics with higher angular resolution.

\section{SUMMARY}
%----------------
\label{sec:conclusion}
%We have presented new X-ray measurements of the two relic galaxies
% 0917+75 and 1401$-$33 with the \emph{XMM-Newton} observatory.
%We detected no IC/CMB flux from either target, but obtained new limits
% on the IC/CMB flux and magnetic field.
With new X-ray measurements of the two relic sources
 0917+75 and 1401$-$33 using the \emph{XMM-Newton} observatory,
 we detected no IC/CMB flux from either target.
%We set the $2\sigma$ upper limits on the IC/CMB flux within 0.3--10\,keV
% at  $1.38 \times 10^{-14}$\,erg~s$^{-1}$~cm$^{-2}$ for 0917+75
% and $9.71 \times 10^{-14}$\,erg~s$^{-1}$~cm$^{-2}$ for 1401$-$33.
%The corresponding $2\sigma$ lower limits on their magnetic field strengths are
% 0.93\,$\mu$G and 1.9\,$\mu$G, respectively,
We set the $3\sigma$ upper limits on the IC/CMB flux within 0.3--10\,keV
 at  $5.22 \times 10^{-14}$\,erg~s$^{-1}$~cm$^{-2}$ for 0917+75
 and $1.47 \times 10^{-13}$\,erg~s$^{-1}$~cm$^{-2}$ for 1401$-$33.
The corresponding $3\sigma$ lower limits on their magnetic field strengths are
 0.81\,$\mu$G and 2.2\,$\mu$G, respectively,
 both slightly larger than the classical estimates of the equipartition field.
\iffalse
% listed in Table~\ref{tab:resultsInferred},
% are inconsistent with the values predicted
% by the equipartition argument: $B$ is twice too high for 0917+75,
% and at least four times too high for 1401$-$33.
%We reviewed what we know about these relics
% from past radio and X-ray studies of the two sources.
%We discussed possible causes of the departure from equipartition;
These fields are slightly higher than the values obtained
 from the standard equipartition formula,
 but there are enough uncertainties that make this physically insignificant.
%Possible causes of the difference between the measured
% magnetic field strengths and the values obtained from the standard
% equipartition formula are:
Possible causes of the difference are:
 (1) a sizable contribution to the particle energy by protons,
 (2) a less-than-unity filling factor for the emitting plasma,
 (3) spectral breaks in the radio spectra at low frequencies
 that are yet unaccessable, and
 (4) genuine deviation from equipartition, with a higher energy density
 in the magnetic field than in particles.
Regardless of the true cause, better determination of the magnetic field
 will help to better understand relic galaxies and radio lobes in general.
Further progress can be made by IC/CMB measurements at higher X-ray energies,
 and by synchrotron measurements at lower radio frequencies.
\fi
Our study adds to the small sample of relics with limits on their IC/CMB emission.
X-ray studies on more relics are needed for a broader picture of this diverse class
 of radio sources.
Further constraints on the particle density and magnetic field strength
 can be made by IC/CMB measurements at higher X-ray energies (above 10\,keV),
 and by synchrotron measurements at lower radio frequencies (below 300\,MHz).

%\acknowledgments
\section*{Acknowledgments}
We acknowledge M.~Markevitch for discussion on \emph{XMM-Newton} background,
 and the anonymous reviewer for many constructive suggestions,
 including the two-step background subtraction method
 and better presentation of the images in this paper.
CMHC thanks Wayne H. Baumgartner for many discussions and advices on SAS,
 and Ryan C. Ogliore for help interpreting \emph{ACE} data.
%\footnote{Note to self: I am currently emailing to A. M. Read
%	about his blank sky files.  So, may have to thank him, too.}
%\footnote{What grant(s) should I acknowledge?
%	HEFT?  PECA for FAH?
%	Peter's NASA grad student fellowship?
%	Anything for DEH?}
%\footnote{Thank the referee?  I am not going to, unless you tell me to.}
%This research was supported by the NASA Space Science
%``Supporting Research and Technology'' (SR\&T) programme
%under Grant Number NAG5-5398.
%FAH was further supported by a Presidential Early Career Award,
%Grant Number NAG5-5322.
We thank the \emph{ACE} SWEPAM instrument team and the \emph{ACE} Science Center
 for providing the \emph{ACE} data.
This research has made use
 of SAOImage DS9, developed by the Smithsonian Astrophysical Observatory,
 and of NASA's Astrophysics Data System.


\begin{thebibliography}{}
\bibitem[\protect\citeauthoryear{Arnaud et al.}{2002}]{Arnaud2002}
	Arnaud, M., Majerowicz, S., Lumb, D., Neumann, D. M., Aghanim, N.,
	Blanchard, A., Boer, M., Burke, D. J., Collins, C. A., Giard, M.,
	Nevalainen, J., Nichol, R. C., Romer, A. K., Sadat, R.
% title = {XMM-Newton observation of the distant (z=0.6) galaxy cluster RX J1120.1+4318}
	2002, A\&A, 390, 27

\bibitem[\protect\citeauthoryear{Bacchi et al.}{2003}]{BacchiFerettiGiovanniniGovoni2003}
	Bacchi M., Feretti L., Giovannini G., Govoni, F.,
% title = {Deep images of cluster radio halos}
	2003, A\&A, 400, 465

\bibitem[\protect\citeauthoryear{Bagchi, Pislar \& Lima Neto}{Bagchi et al.}{1998}]{1998MNRAS.296L..23B}
	Bagchi J., Pislar V., Lima Neto G. B.,
%    title = {The diffuse, relic radio source in Abell 85: estimation of cluster-scale magnetic field from inverse Compton X-rays},
	1998, MNRAS, 296, L23

\bibitem[\protect\citeauthoryear{Beck \& Krause}{2005}]{BeckKrause2005}
	Beck R., Krause M.,
% title = {Revised equipartition and minimum energy formula for magnetic field strength estimates from radio synchrotron observations}
	2005, Astron. Nachr., 326, 414

\bibitem[\protect\citeauthoryear{Brunetti, Setti \& Comastri}{Brunetti et al.}{1997}]{BrunettiSettiComastri1997}
	Brunetti G., Setti G., Comastri A.,
% title = {Inverse Compton X-rays from strong FRII radio-galaxies}
	1997, A\&A, 325, 898

%\bibitem[Carilli \& Taylor(2002)]{CarilliTaylor2002}
%	Carilli, C. L., \& Taylor, G. B.
%% title = {Cluster Magnetic Fields}
%	2002, ARA\&A, 40, 319

\bibitem[\protect\citeauthoryear{Carter \& Read}{2007}]{CarterRead2007}
	Carter J. A., Read A. M.,
%% title = {The XMM-Newton EPIC background and the production of background blank sky event files}
	2007, A\&A, 464, 1155

\bibitem[\protect\citeauthoryear{Condon et al.}{1998}]{nvss}
	Condon J. J., Cotton W. D., Greisen E. W., Yin Q. F.,
	Perley R. A., Taylor G. B., Broderick J. J.,
	1998, AJ, 115, 1693

%\bibitem[Cordey(1987)]{Cordey1987}
%	Cordey, R. A.
%%  title    = {IC-2476 - A POSSIBLE RELIC RADIO GALAXY},
%	1987, MNRAS, 227, 695

\bibitem[\protect\citeauthoryear{Croston et al.}{2005}]{Croston2005}
	Croston J. H., Hardcastle M. J., Harris D. E., Belsole E.,
	Birkinshaw M., Worrall D. M.,
%    title = {An X-Ray Study of Magnetic Field Strengths and Particle Content in the Lobes of FR II Radio Sources}
	2005, ApJ, 626, 733

\bibitem[\protect\citeauthoryear{Dewdney et al.}{1991}]{DewdneyCostainMchardy1991}
	Dewdney P. E., Costain C. H., McHardy I.,
		Willis A. G., Harris D. E., Stern C. P.,
%  title    = {AN X-RAY AND RADIO STUDY OF STEEP-SPECTRUM RADIO-
%              SOURCES .2. 4 FIELDS FROM A 22 MHZ POLAR-CAP SURVEY},
	1991, ApJS, 76, 1055

\bibitem[\protect\citeauthoryear{Durret, Lima Neto \& Forman}{Durret et al.}{2005}]{DurretLimaNetoForman2005}
	Durret F., Lima Neto G. B., Forman W.,
% title = {An XMM-Newton view of the cluster of galaxies Abell 85}
	2005, A\&A, 432, 809

%\bibitem[En{\ss}lin et al.(1998)]{Ensslin1998}
%	En{\ss}lin, T. A., Biermann, P. L., Klein, U., \& Kohle, S.
%% title = {Cluster radio relics as a tracer of shock waves of the large-scale structure formation}
%	1998, A\&A, 332, 395

%\bibitem[Feigelson et al.(1995)]{1995ApJ...449L.149F}
%	Feigelson, E. D., Laurent-Muehleisen, S. A., Kollgaard, R. I.,
%	\& Fomalont, E. B.
%%    title = {Discovery of Inverse-Compton X-Rays in Radio Lobes},
%	1995, ApJ, 449, L149

\bibitem[\protect\citeauthoryear{Feretti \& Giovannini}{1996}]{FerettiGiovannini1996}
	Feretti L., Giovannini G.,
% title = {Diffuse Cluster Radio Sources (Review)}
	1996, in Ekers R. D., Fanti C., Padrielli L., eds,
	Proc. IAU Symp. 175, Extragalactic Radio Sources.
	Kluwer Academic Publishers, p.~333

\bibitem[\protect\citeauthoryear{Feretti \& Neumann}{2006}]{FerettiNeumann2006}
	Feretti L., Neumann D. M.,
% title = {XMM-Newton observations of the Coma cluster relic 1253+275}
	2006, A\&A, 450, L21

\bibitem[\protect\citeauthoryear{Fusco-Femiano et al.}{2003}]{Fusco-Femiano2003}
	Fusco-Femiano R., Orlandini M., De Grandi S., Molendi S.,
	Feretti L., Giovannini G., Bacchi M., Govoni F.,
%  title    = {Hard X-ray and radio observations of Abell 754}
	2003, A\&A, 398, 441

\bibitem[\protect\citeauthoryear{Fujita et al.}{2004}]{Fujita2004}
	Fujita Y., Sarazin C. L., Reiprich T. H., Andernach H.,
	Ehle M., Murgia M., Rudnick L., Slee O. B.,
%    title = {XMM-Newton Observations of A133: A Weak Shock Passing through the Cool Core}
	2004, ApJ, 616, 157

\bibitem[\protect\citeauthoryear{Gavazzi \& Trinchieri}{1983}]{GavazziTrinchieri1983}
	Gavazzi G., Trinchieri G.,
%    title = {Radio and X-ray observations of the radio halo source in A1367}
	1983, ApJ, 270, 410

\bibitem[\protect\citeauthoryear{Giovannini \& Feretti}{2004}]{GiovanniniFeretti2004}
	Giovannini G., Feretti L.,
% title = {Radio relics in clusters of galaxies}
	2004, J. Kor. Astron. Soc., 37, 323

%\bibitem[Giovannini et al.(1991)Giovannini, Feretti, \& Stanghellini]{GiovanniniFerettiStanghellini1991}
%	Giovannini, G., Feretti, L., \& Stanghellini, C.
%%  title    = {THE COMA CLUSTER RADIO-SOURCE 1253+275, REVISITED},
%	1991, A\&A, 252, 528

\bibitem[\protect\citeauthoryear{Giovannini, Tordi \& Feretti}{Giovannini et al.}{1999}]{GiovanniniTordiFeretti1999}
	Giovannini G., Tordi M., Feretti L.,
%  title    = {Radio halo and relic candidates from the NRAO VLA Sky Survey}
	1999, New Astron., 4, 141

\bibitem[\protect\citeauthoryear{Goss et al.}{1987}]{GossMcadamWellington1987}
	Goss W. M., McAdam W. B., Wellington K. J., Ekers R. D.,
%  title    = {THE VERY LOW-BRIGHTNESS RELIC RADIO GALAXY 1401-33},
	1987, MNRAS, 226, 979

\bibitem[\protect\citeauthoryear{Harris \& Grindlay}{1979}]{1979MNRAS.188...25H}
	Harris D. E., Grindlay J. E.,
%    title = {The prospects for X-ray detection of inverse-Compton emission from radio source electrons and photons of the microwave background},
	1979, MNRAS, 188, 25

\bibitem[\protect\citeauthoryear{Harris et al.}{1993}]{HarrisSternWillis1993}
	Harris D. E., Stern C. P., Willis A. G., Dewdney P. E.,
%  title    = {The megaparsec radio relic in supercluster, Rood number 27},
	1993, AJ, 105, 769

\bibitem[\protect\citeauthoryear{Harris et al.}{1995}]{HarrisWillisDewdney1995}
	Harris D. E., Willis A. G., Dewdney P. E., Batty J.,
%  title    = {Constraints on the average magnetic-field strength in a relic
%              radio galaxy derived from limits on inverse Compton X-rays},
	1995, MNRAS, 273, 785

%\bibitem[Henriksen(1998)]{1998PASJ...50..389H}
%	Henriksen, M.
%%    title = {X-Ray Constraints on Cluster Magnetic Fields},
%	1998, PASJ, 50, 389

\bibitem[\protect\citeauthoryear{Henriksen, Hudson \& Tittley}{Henriksen et al.}{2004}]{HenriksenHudsonTittley2004}
	Henriksen M., Hudson D. S., Tittley E.,
%    title = {A soft X-ray excess in the A754 cluster}
	2004, ApJ, 610, 762

\bibitem[\protect\citeauthoryear{Henriksen \& Mushotzky}{2001}]{HenriksenMushotzky2001}
	Henriksen M., Mushotzky R.,
%    title = {X-ray measurements of the nonthermal emission from the Abell 1367 galaxy cluster using the ROSSI X-ray Timing Explorer}
	2001, ApJ, 553, 84

%\bibitem[Jamrozy et al.(2004)]{Jamrozy2004}
%	Jamrozy, M., Klein, U., Mack, K.-H., Gregorini, L., \& Parma, P.
%%title = {Spectral ageing in the relic radio galaxy B2 0924+30}
%	2004, A\&A, 427, 79

%\bibitem[Kaneda et al.(1995)]{1995ApJ...453L..13K}
%%	Kaneda, H., Tashiro, M., Ikebe, Y., Ishisaki, Y., Kubo, H.,
%%	Makshima, K., Ohashi, T., Saito, Y., Tabara, H., \& Takahashi, T.
%	Kaneda, H., et al.
%%    title = {Detection of Inverse-Compton X-Rays from Lobes of the Radio Galaxy Fornax A},
%	1995, ApJ, 453, L13

\bibitem[\protect\citeauthoryear{Kempner et al.}{2004}]{Kempner2004}
	Kempner J. C., Blanton E. L., Clarke T. E.,
	En{\ss}lin T. A., Johnston-Hollitt M., Rudnick L.,
%    title = {Conference Note: A Taxonomy of Extended Radio Sources in Clusters of Galaxies}
	2004, in Reiprich T. H., Kempner J. C., Soker N., eds,
	The Riddle of Cooling Flows in Galaxies and Clusters of Galaxies.
	p.~E25

\bibitem[\protect\citeauthoryear{Kempner \& Sarazin}{2001}]{KempnerSarazin2001}
	Kempner J. C., Sarazin C. L.,
%    title = {Radio Halo and Relic Candidates from the Westerbork Northern Sky Survey}
	2001, ApJ, 548, 639

\bibitem[\protect\citeauthoryear{Kempner, Sarazin \& Markevitch}{Kempner et al.}{2003}]{KempnerSarazinMarkevitch2003}
	Kempner J. C., Sarazin C. L., Markevitch M.,
%    title = {Chandra observation of the merging cluster A2034}
	2003, ApJ, 593, 291

%\bibitem[\protect\citeauthoryear{Leahy \& Gizani}{2001}]{LeahyGizani2001}
%	Leahy J. P., Gizani N. A. B.,
%%    title = {ROSAT Observations of 3C 388: A Test of Minimum Energy}
%	2001, ApJ, 555, 709

\bibitem[\protect\citeauthoryear{McGlynn, Scollick \& White}{McGlynn et al.}{1996}]{skyview}
	McGlynn T., Scollick K., White N.,
% title = {SkyView: The Multi-Wavelength Sky on the Internet}
	1996, in McLean B. J., Payne H. E., eds,
	Proc. IAU Symp. 179, New Horizons from Multi-Wavelength Sky Surveys.
	Kluwer Academic Publishers, Boston, p.~465

\bibitem[\protect\citeauthoryear{Mukai}{1993}]{pimms}
	Mukai K.,
	1993, Legacy, 3, 21

\bibitem[\protect\citeauthoryear{Nevalainen, Markevitch \& Lumb}{Nevalainen et al.}{2005}]{Nevalainen}
	Nevalainen J., Markevitch M., Lumb D.,
%    title = {XMM-Newton EPIC background modeling for extended sources},
	2005, ApJ, 629, 172

%\bibitem[Newman et al.(2002)Newman, Newman, \& Rephaeli]{NewmanNewmanRephaeli2002}
%	Newman, W. I., Newman, A. L., \& Rephaeli, Y.
%%    title = {Quantification of uncertainty in the measurement of magnetic fields in clusters of galaxies}
%	2002, ApJ, 575, 755

\bibitem[\protect\citeauthoryear{Pacholczyk}{1970}]{Pacholczyk1970}
	Pacholczyk A. G.,
	1970, Radio Astrophysics.  W. H. Freeman and Company, San Francisco, CA

\bibitem[\protect\citeauthoryear{Pfrommer \& En{\ss}lin}{2004}]{PfrommerEnsslin2004}
	Pfrommer C., En{\ss}lin T. A.,
%  title    = {Estimating galaxy cluster magnetic fields by the classical and hadronic minimum energy criterion}
	2004, MNRAS, 352, 76

\bibitem[\protect\citeauthoryear{Pradas \& Kerp}{2005}]{PradasKerp2005}
	Pradas J., Kerp J.,
% title = {XMM-Newton data processing for faint diffuse emission}
	2005, A\&A, 443, 721

%\bibitem[Read \& Ponman(2003)]{ReadPonman2003}
%	Read, A. M., \& Ponman, T. J.
%% title = {The XMM-Newton EPIC background: Production of background maps and event files}
%	2003, A\&A, 409, 395

\bibitem[\protect\citeauthoryear{Slee et al.}{2001}]{Slee2001}
	Slee O. B., Roy A. L., Murgia M., Andernach H., Ehle M.,
% title = {Four extreme relic radio sources in clusters of galaxies}
	2001, AJ, 122, 1172

\bibitem[\protect\citeauthoryear{Snowden, Collier \& Kuntz}{Snowden et al.}{2004}]{Snowden}
	Snowden S. L., Collier M. R., Kuntz K. D.,
% title = {XMM-Newton Observation of Solar Wind Charge Exchange Emission}
	2004, ApJ, 610, 1182

\bibitem[\protect\citeauthoryear{Subrahmanyan et al.}{2003}]{Subrahmanyan2003}
	Subrahmanyan R., Beasley A. J., Goss W. M., Golap K., Hunstead R. W.,
% title = {PKS B1400-33: An Unusual Radio Relic in a Poor Cluster}
	2003, AJ, 125, 1095

%\bibitem[\protect\citeauthoryear{Tashiro, Makishima \& Kaneda}{Tashiro et al.}{2000}]{TashiroMakishimaKaneda2000}
%	Tashiro M., Makishima K., Kaneda H.,
%% title = {ASCA Measurements of Field-Particle Energy Distribution in Radio Lobes}
%	2000, Adv. Space Res., 25, 751

%\bibitem[Tashiro et al.(1998)]{1998ApJ...499..713T}
%%	Tashiro, M., Kaneda, H., Makishima, K., Iyomoto, N.,
%%        Idesawa, E., Ishisaki, Y., Kotani, T., Takahashi, T., \& Yamashita, A.
%	Tashiro, M., et al.
%%    title = {Evidence of Energy Nonequipartition between Particles and Fields in Lobes of the Radio Galaxy PKS 1343-601 (Centaurus B)},
%	1998, ApJ, 499, 713

%\bibitem[\protect\citeauthoryear{Valtchanov et al.}{2002}]{Valtchanov2002}
%	Valtchanov I., Murphy T., Pierre M., Hunstead R., L\'{e}monon L.,
%%title = {Abell 1451 and 1RXS J131423.6-251521: A multi-wavelength study of two dynamically perturbed clusters of galaxies}
%	2002, A\&A, 392, 795

%\bibitem[van der Laan \& Perola (1969)]{vanderLaanPerola1969}
%	van der Laan, H., \& Perola, G. C.
%%    title = {Aspects of Radio Galaxy Evolution},
%	1969, A\&A, 3, 468
\end{thebibliography}
\end{document}